\documentclass[aps,prb,twocolumn,superscriptaddress,longbibliography]{revtex4-2}
\usepackage{graphicx}
\usepackage{amsmath}
\usepackage{bbold}
\usepackage[colorlinks=true, citecolor=blue, urlcolor=blue, linkcolor=blue,bookmarks=false,hypertexnames=true]{hyperref} 
\usepackage[normalem]{ulem}
\usepackage{xcolor}

\def \DTP {Department of Theoretical Physics, Wroc\l{}aw University of Science and Technology, Wybrze\.{z}e Wyspia\'{n}skiego 27, 50-370 Wroc\l{}aw, Poland}

\def \DECE {Department of Electrical and Computer Engineering, University of Illinois at Chicago, Chicago, IL 60607, USA}

\begin{document}
	
\title{Electrical manipulation of valley-qubit and valley geometric phase\\ in lateral monolayer heterostructures}

\author{Jaros\l{}aw Paw\l{}owski}\email[]{jaroslaw.pawlowski@pwr.edu.pl}\affiliation{\DTP}
\author{John Eric Tiessen}\affiliation{\DECE}
\author{Rockwell Dax}\affiliation{\DECE}
\author{Junxia Shi}\affiliation{\DECE}

\begin{abstract}
We explore a solid state qubit defined on valley isospin of an electron confined in a gate-defined quantum dot created in an area of monolayer MoS$_2$/WS$_2$ lateral junction,
where a steep dipolar potential emerges. We show that the junction oriented along an armchair direction can induce intervalley transitions of the electron confined in the neighboring quantum dot when the (gate-controllable) overlapping with the junction is significant and pumping frequency tuned.
The pumping scheme that induces transitions is all-electrical: obtained by applying oscillating voltages to control gates and thus enables for scalable qubit architectures.
We also report another possibility of valley-qubit manipulation by accumulating non-Abelian valley Berry phase.
To model nanodevice we solve the time-dependent Schr\"odinger-Poisson equations in a tight-binding approach and obtain exact time-evolution of the valley-qubit system. 
\end{abstract}

\maketitle

\section{Introduction}
Advances continue to be made in the implementation of quantum computing hardware. However, much of the recent progress in quantum computing chips has been made with superconducting qubits \cite{superconducting_current_state}. 
Such implementations of quantum computing hardware have so far been limited to dozens of qubits rather than the thousands needed for fault tolerant quantum computing architectures \cite{science_material_challenges}. Therefore, alternative types of qubits, as well as different qubit host materials, have been an area of great interest in the last several years \cite{programmable_two_qubit, QD_array_Si_Ge, two_axis_control}. Most of these proposed qubit systems seek to use silicon as the host material since it is well understood, mainly due to its long history of use in the semiconductor industry. However, silicon has intrinsic limitations, such as
relatively weak spin-orbit coupling, or isotopic impurities leading to spin dechoherence,
which also restrict its utility as a qubit host. 
These issues can be remedied~\cite{cross_bar_network, enhanced_electrostatic_coupling}, 
nonetheless, a more suitable qubit host material should be found.

2D materials such as Transition Metal Dichalcogenides (TMDCs) and Bi-Layer Graphene (BLG) offer an alternative approach to the problem of manipulating and isolating qubits. TMDCs in particular are an excellent candidate for qubit hosting and manipulation~\cite{Burkard2014,Fabian2021,mymos2,myrash,Brooks2020,Pawlowski2021,menaf,justin2021} due to their intrinsic spin-orbit interaction and 2D nature. In this work we propose a 2D lateral TMDC heterostructure~\cite{oscar2} to act as a qubit host material, while using the position dependent change in the electronic properties of the heterostructure near the interface to control the state of the qubit. Due to recent progress in the growth of lateral TMDC heterostructures~\cite{recent_hetero, small_structures}, such a structure seems reasonable from an experimental/engineering perspective~\cite{oscar2,satwik}. Also, the all-electric nature of the proposed qubit means that the footprint needed for device implementation can be brought down to the order of nanometers. 

Examples of 2D lateral TMDC heterostructures include the MoS(Se)$_2$/WS(Se)$_2$ semiconductor
heterostructures~\cite{tmd-lateral1,tmd-lateral2,tmd-lateral3,tmd-lateral4,tmd-lateral5,wse2_strain}. 
In particular, MoS$_2$/WS$_2$~\cite{tmd-lateral2} and MoS$_2$/WSe$_2$~\cite{tmd-lateral4, wse2_strain} heterostructures with an atomically sharp interface in the armchair or zigzag configuration have been successfully synthesized. 
Also interesting physical properties for device applications using TMDC lateral junctions, such as the photovoltaic effect, have been demonstrated experimentally~\cite{recent_hetero,oscar2}.
Such structures also open the way to develop ultra-efficient, planar thermoelectric devices~\cite{thermo}.
An important limitation, however, is that lateral interfaces can only be realized in epitaxially grown TMDCs, which are known to exhibit lower quality than mechanically exfoliated crystals.

In addition to the spin degree of freedom, TMDCs also possess a valley isospin. This additional degree provides an extra set of states for defining qubits, but the problem arises of how to efficiently manipulate such qubit states. To achieve this, one may use optical manipulation~\cite{Yue2016}, but for a scalable architecture electric control is desired. To control the qubit via local gating, creation of a confinement potential with a size comparable to the lattice vector is needed~\cite{Liu_2014, Sz_chenyi_2018, myrash, menaf}, which is difficult to engineer using presently available gate lithography resolutions. 

In our proposed system, this problem is overcome via the utilization of the sharp potential profile naturally generated at the lateral heterojunction of the two 2D materials. 
Such an inter-TMDC-monolayer interface (assuming that it is atomically sharp and of armchair type) generates a steep dipolar potential that changes the momentum profile of the gate-induced confining potential for the electron, which enables the coupling between the states in $K$ and $K'$ valleys. The magnitude of this effect depends on the distance between the electron and the junction that can be varied by changing the applied gate voltage.

The paper is organized as follows: In Section~\ref{sec:model} we will introduce the device structure based on the TMDC heterojunction, as well as the theoretical model which describes spin-valley physics of a single-electron qubit carrier confined within the junction area. In the next Section~\ref{sec:reslt} we present results on valley qubit manipulation, while in Section~\ref{sec:cond} we analyze conditions that should be met to couple valley states. In the final Section,~\ref{sec:geom}, we study the possibility of valley manipulation by  geometric Berry phase accumulation.

\section{Device model}\label{sec:model}
\begin{figure}[t]
	\center
	\includegraphics[width=7.5 cm]{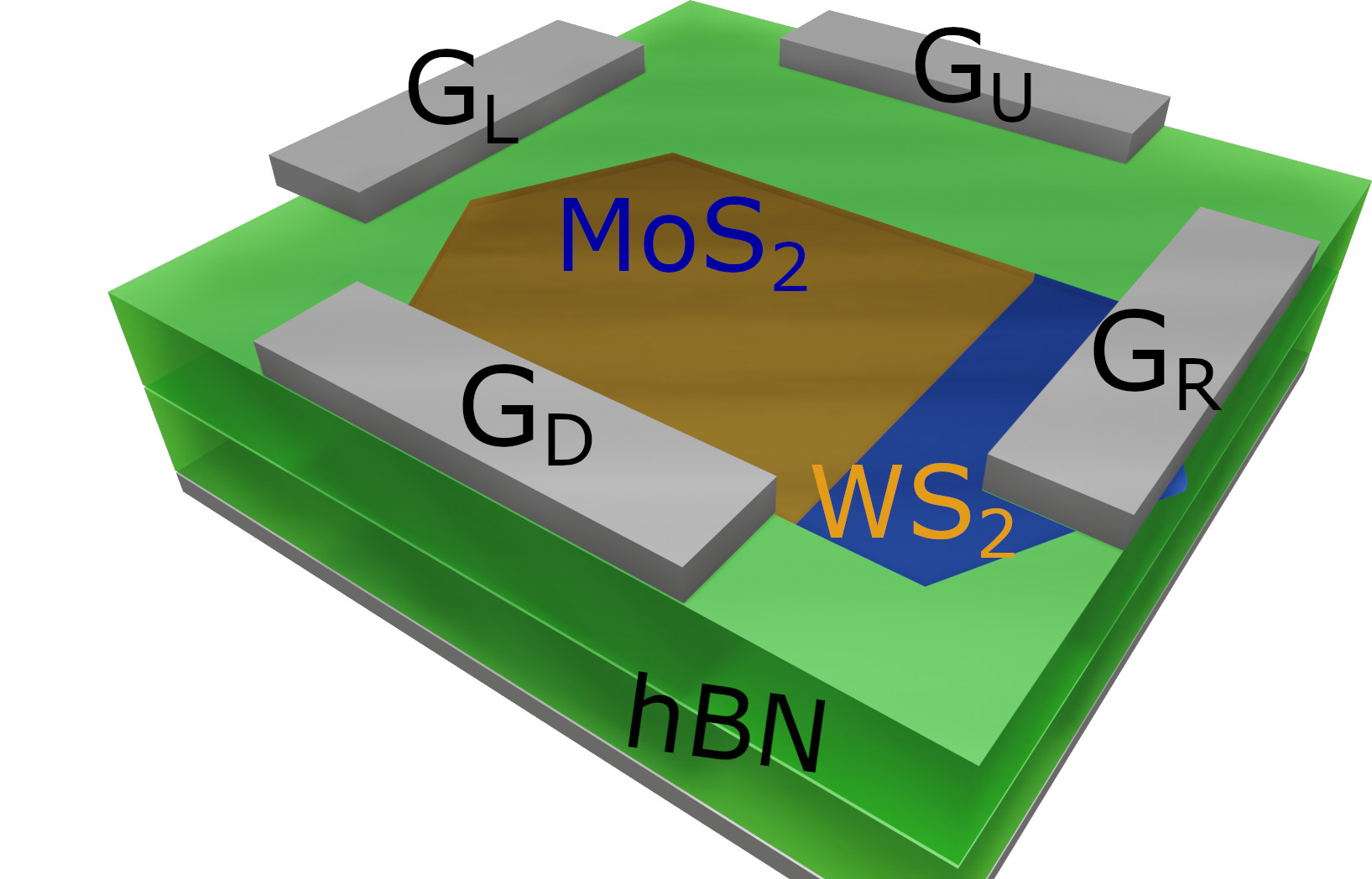}
	\caption{\label{fig:1} The schematic view on the proposed device structure containing MoS$_2$/WS$_2$ heterojunction (orange/blue) sandwiched between hBN tunnel barriers, together with four gates $\mathrm{G}_\mathrm{\{L,R,U,D\}}$ deposited on top defining quantum dot in the junction area.}
\end{figure}
\begin{figure}[b]
	\center
	\includegraphics[width=8.8cm]{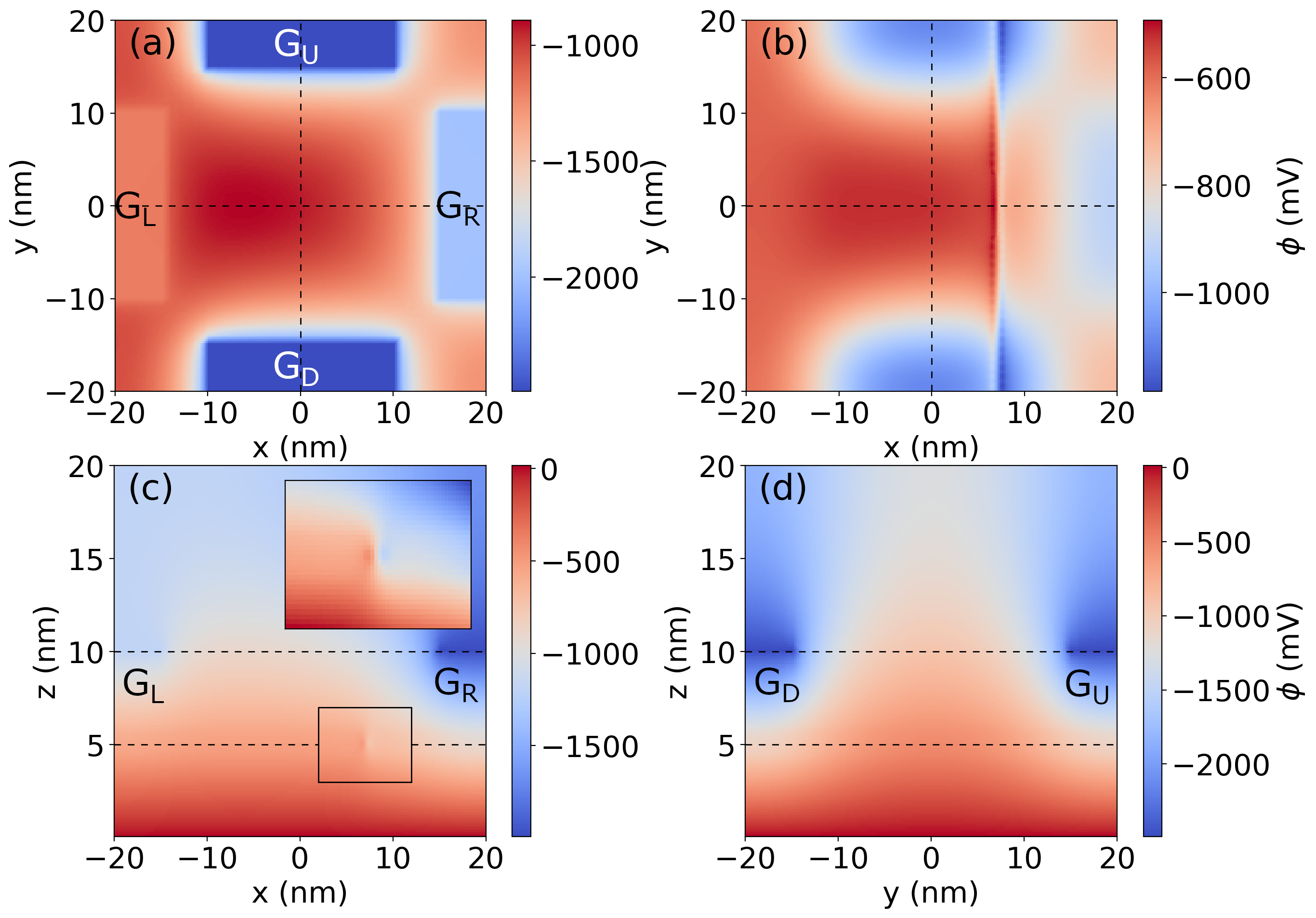}
	\caption{\label{fig:2} Electrostatic potential profiles within the nanodevice. Cross-sections: (a) at the top gates level (with marked gates positions), (b) monolayers level with visible junction profile, or along (c) $y=0$ surface with visible junction dipole (inset), and (d) $x=0$ surface.} 
\end{figure}

The proposed nanodevice is composed of a MoS$_2$/WS$_2$ lateral (in-plane) heterojunction sandwiched between hBN tunnel barriers each $5$~nm thick -- see Fig.~\ref{fig:1}. On top of the upper hBN layer four metallic gates are deposited, $\mathrm{G}_\mathrm{\{L,R,U,D\}}$. Their role is to create a quantum dot (QD) confinement potential in the junction area, and to control valley coupling (via $\mathrm{G}_\mathrm{L}$ and $\mathrm{G}_\mathrm{R}$) by tuning the confined electron's wavefunction overlap with the junction dipole potential profile. The whole structure is placed on the highly doped substrate which acts as a backgate with zero reference voltage. Profiles of the QD potential defined by the top gates, and calculated using the Poisson-Schr\"odinger method, are presented in Fig.~\ref{fig:2}(a) -- cross-section at four top gates level, $z_g=10$~nm, and Fig.~\ref{fig:2}(b) -- cross-section at monolayers level, $z_l=5$~nm with also visible junction potential profile at $x_j=7$~nm.
In Fig.~\ref{fig:2}(c) with presented lateral $x$-$z$ cross-section through the device we can observe electric dipole at the junction (zoomed inset) and constant potential conditions by applying voltages to the $\mathrm{G}_\mathrm{L}$ and $\mathrm{G}_\mathrm{R}$ top gates. Also top gates $\mathrm{G}_\mathrm{U}$ and $\mathrm{G}_\mathrm{D}$ are visible but in cross-section in $y$-$z$ directions presented in Fig.~\ref{fig:2}(d).
It should also be added that in realistic configurations, the thickness of the hBN spacer can be increased, e.g. to 10-15~nm, to avoid voltage breakdown, but this may slightly reduce the electron confinement.

\subsection{Heterojunction}
To realistically model the junction between two 2D monolayer materials we have made the following assumptions. An MoS$_2$/WS$_2$ junction possess type-II band alignment~\cite{Lee_2017} with a conduction band (CB) offset of approximately $\sim0.35$~eV~\cite{junction1,junction2}. This can be calculated using the Anderson rule and the difference in electron affinity of these two materials. 
In addition to the band offset caused by connecting the two semiconductors, the Fermi levels of the two semiconductors must also match. This matching leads to characteristic band bending in the junction area and determines the exact shape of the junction potential profile.
Fermi level misalignment generates charge flow
until $E_F$ equilibrates creating space charges (charge transfer across the interface~\cite{cht1,cht2} from MoS$_2$ to WS$_2$) that build up forming the electrostatic dipole at the junction~\cite{junction2}. This induced dipole is strongly localized to the heterojunction. Therefore, to model this effect we assume two charge densities which form a dipole linear charge density with a value of $\lambda=378\,|e|/\mathrm{nm}$.
Next, we assume the following density distribution for space charge:
\begin{equation}\label{eq:charge}
    d^\pm(x,y,z)=\frac{\pm\lambda}{2\pi\sigma_x\sigma_z}e^{-\frac{(x-x_j\pm \delta)^2}{2\sigma_x^2}-\frac{(z-z_l)^2}{2\sigma_z^2}},
\end{equation}
with $\sigma_x=\sigma_z=0.1$~nm, and charge displacement $\delta=0.25$~nm.
The linear density $\lambda$ parameter was tuned in our model to give the desired CB offset $0.35$~eV. Its actual value is about 3 times larger than the one derived from [\onlinecite{junction3}]: $\lambda=130\,|e|/\mathrm{nm}$.
We take $d^+(x,y,z)$ at MoS$_2$ side, $d^-(x,y,z)$ at WS$_2$ side and 0 elsewhere in computational box (i.e. outside the monolayers). 

\begin{figure}[t]
	\center
	\includegraphics[width=8.8cm]{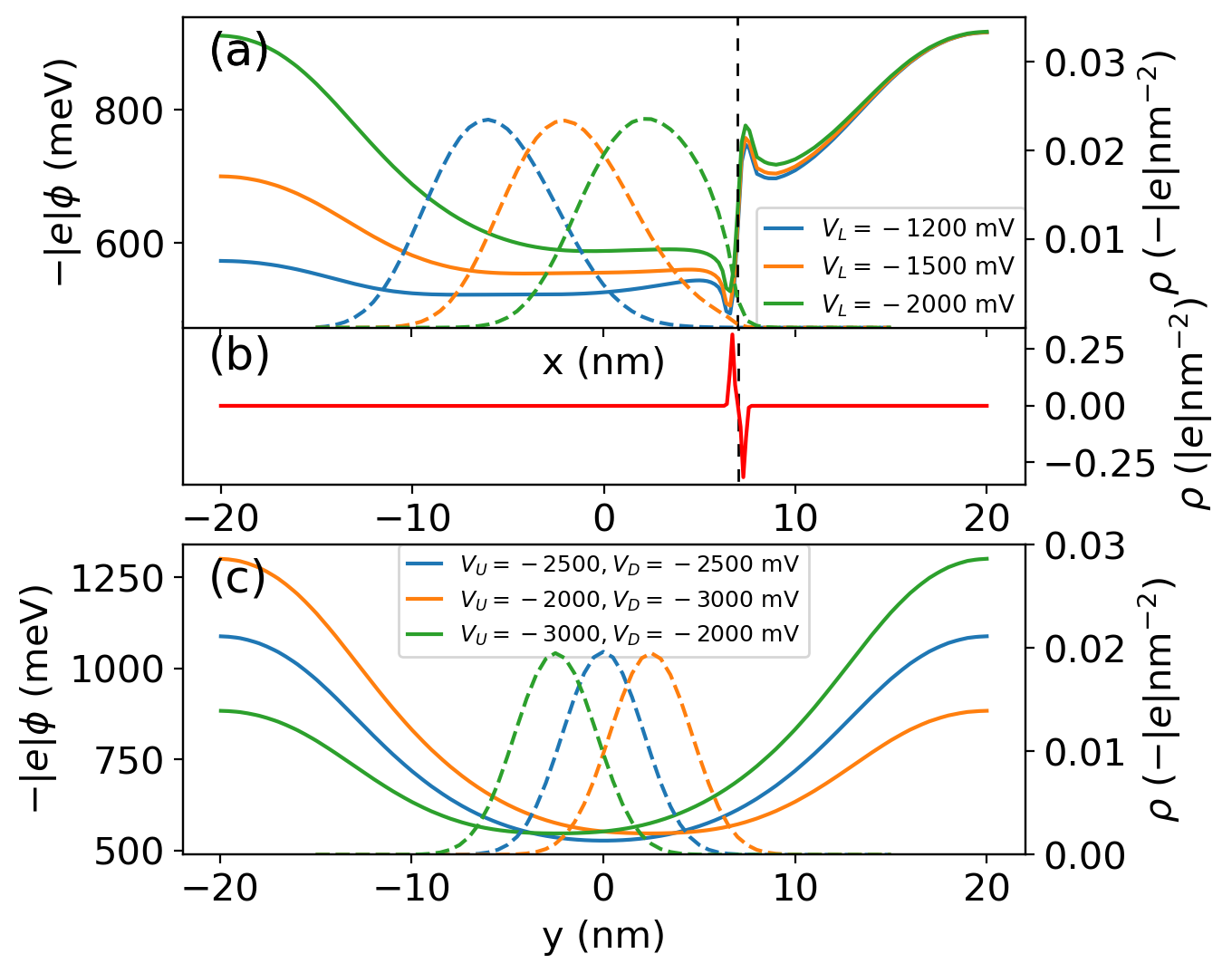}
	\caption{\label{fig:3} (a) Potential profiles (solid lines) through the junction interface, with the characteristic CB offset built by space charge dipole (b). Modulation of top gate voltages changes the potential, also in direction along the junction (c), and shifts electron (dashed lines) confined within the QD.}
\end{figure}

The resulting spatial charge dipole density is presented in Fig.~\ref{fig:3}(b). This dipole generates a built-in electric field across the interface as seen in Fig.~\ref{fig:3}(a) (solid lines). The obtained CB offset is of the desired value.
Also it should be noted that we assume that the difference in electron affinities and work functions are similar and therefore the potential levels on both sides far from the junction match with each other. 
If we additionally modulate the voltage applied to the left gate $\mathrm{G}_\mathrm{L}$ we may tune the shape of the confinement potential (mostly at the MoS$_2$ side) which controls the position of the electron density (dashed lines) confined in the QD. In this way by changing $V_\mathrm{L}$ we may control the amount of the electron density which is localized at the junction. A similar control scheme may by done in the $y$ direction by instead using the voltages $V_\mathrm{U}$ and $V_\mathrm{D}$ -- see Fig.~\ref{fig:3}(c).

Sulphur vacancies are found to be the dominant defect in MoS$_2$ and WS$_2$,~\cite{junction2}. The implemented junction model assumes pristine monolayer materials, however in the case of defects/dopants present in monolayers which change at the Fermi levels, additional source-drain bias will be needed to restore the desired junction profile. 

\subsection{Tight-binding model}

\begin{figure*}[t]
	\center
	\includegraphics[width=16cm]{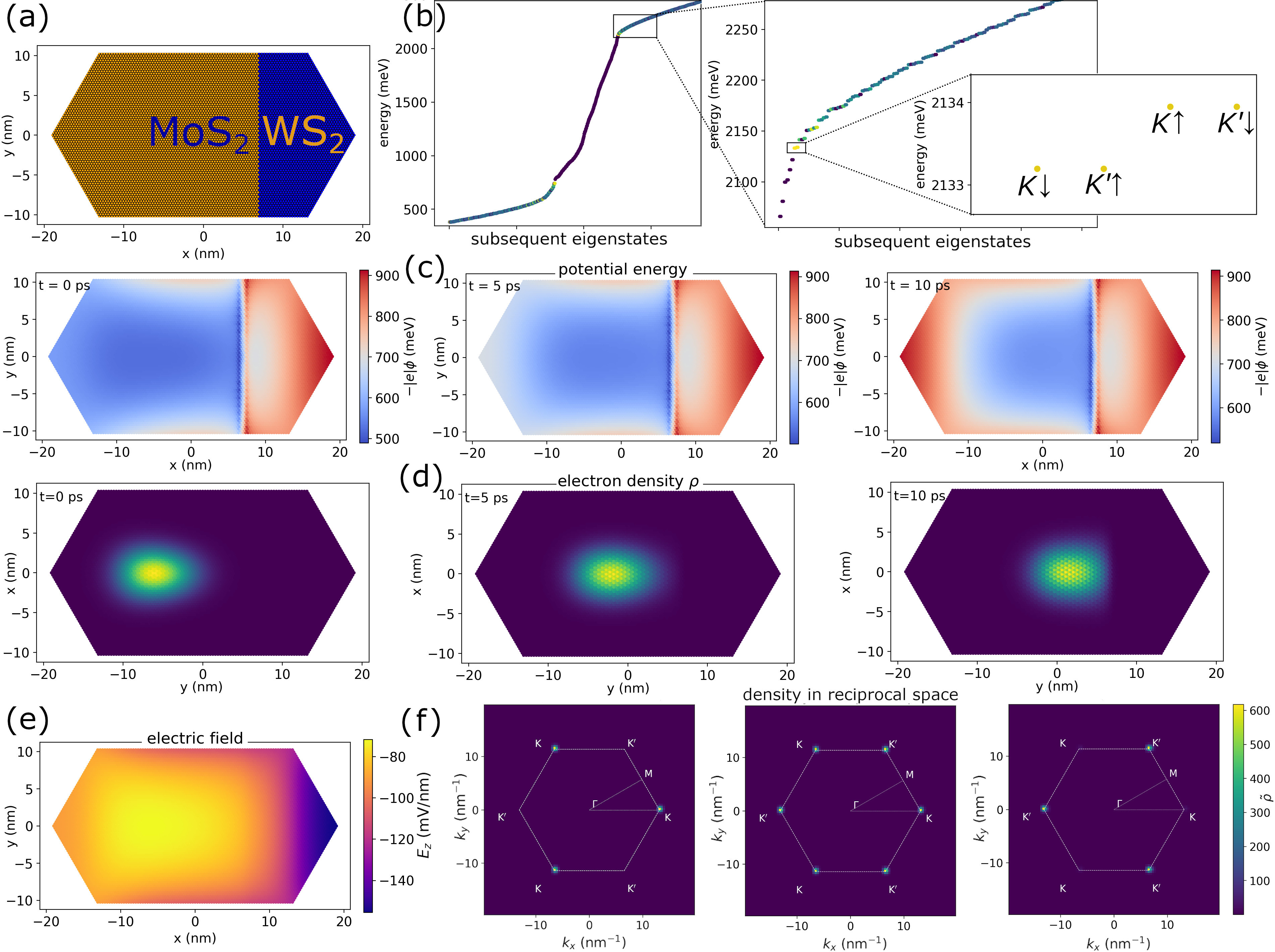}
	\caption{\label{fig:4} (a) Studied nanoflake model with lateral heterostructure (located at $x_j=7$~nm) of two materials. (b) quantum dot eigenstates (colors denote localization of states in QD -- yellow states strongly localize, while dark violet are in-gap edge states) with the lowest CB states (inset) forming spin-valley subspace from which we will select (upon applying B field) two states ($|K\!\downarrow\rangle$, $|K'\!\downarrow\rangle$) that form valley qubit. (c) Confinement potential profile within the fake lattice at initial step and after some confinement modulation that push the electron packet toward the junction barrier -- visible in (d). (e) Perpendicular electric field that induce Rashba SOC. (f) Resonant confinement potential modulation induces intervalley transition gradually transforming density from $K$ to $K'$ valley.
	}
\end{figure*}

To model a single-electron confined in the in-plane monolayer junction area, we develop an atomistic tight-binding (TB) model with different parameters for both sides of the heterostructure and average the hopping parameter through the interface~\cite{oscar1,oscar2}:
\begin{align}\label{ham1e}
H=&\,\sum_{\substack{m\in A(B)\\\alpha\,\sigma}}(\epsilon^{A(B)}_{\alpha }+{\varphi }_m)\,\hat{c}_{m\alpha\sigma}^{\dagger}\hat{c}_{m\alpha\sigma}\nonumber\\
+&\sum_{\substack{m\in A(B)\\\alpha \beta\,\sigma \sigma'}}{s}^z_{\sigma \sigma'}{\lambda^{A(B)} }_{\alpha \beta }\,{\hat{c}}^{\dagger }_{m\alpha \sigma }{\hat{c}}_{m\beta \sigma '}+H_Z+H_R\nonumber\\
+&\sum_{\substack{\langle m, n\rangle\in A(B)\\\alpha \beta\,\sigma}}{t_{\alpha \beta }^{A(B)\langle mn\rangle}\,{\hat{c}}^{\dagger }_{m\alpha \sigma }{\hat{c}}_{n\beta \sigma }}\nonumber\\
+&\sum_{\substack{\langle m\in A, n\in B\rangle\\\alpha \beta\,\sigma}}{\gamma\left(t_{\alpha \beta }^{A\langle mn\rangle}+t_{\alpha \beta }^{B\langle mn\rangle}\right){\hat{c}}^{\dagger }_{m\alpha \sigma }{\hat{c}}_{n\beta \sigma}}
+H.c.
\end{align}
Indices $\{m,n\}$, $\{\sigma, \sigma'\}$, and $\{\alpha,\beta\}$ enumerate lattice sites, spins, and orbitals; e.g., operator ${\hat{c}}^{\dagger }_{m\alpha \sigma }$ (${\hat{c}}_{m\alpha \sigma }$) creates (annihilates) an electron with orbital $\alpha$ and spin $\sigma$ at $m$-th lattice site.
Since both materials have similar lattice constants, we assume the same value $a=0.319$~nm on both sides. 
The lattice combined of two materials MoS$_2$ (A) and WS$_2$ (B) is presented in Fig.~\ref{fig:4}(a).
The potential energy of the electrostatic confinement at the $m$-th lattice site $\varphi_m=-|e|\phi(x_m,y_m)$ together with the on-site energies $\epsilon_\alpha^{A(B)}$ enter the diagonal matrix elements. $\lambda_{\alpha\beta}^{A(B)}$ express the intrinsic spin-orbit coupling~\cite{kos} and $s^z$ stands for the $z$-Pauli-matrix. Both of them take different values at $A$($B$) material sides. $H_Z$ stands for the Zeeman Hamiltonian, and $H_R$ for the Rashba spin-orbit term~\cite{myrash}. The hopping elements $t_{\alpha \beta }^{A(B)\langle mn\rangle}$ depend on the nearest neighbours $\langle mn\rangle$ link direction and if the hopping is inside the given material we assume its hopping parameter $A$ or $B$ (second to last element of Eq.~\ref{ham1e}). The situation is quite different for hopping between materials; in such cases we take the simple average of hoppings when crossing the junction, i.e., $\gamma=\frac{1}{2}$~\cite{half} in the last element of Eq.~\ref{ham1e}.

We utilize a TB model which uses only three $d$ metal-orbitals: $d_{z^2}$, $d_{xy}$, $d_{x^2-y^2}$ on a triangular lattice of Mo (or W atoms), with the nearest-neighbors hoppings \cite{xiao}.
This simple model can correctly represent the dispersion relation and the orbital composition close to the $K$ point in the Brillouin zone (BZ) near the band edges, where the Bloch states mainly consist of metal $d$ orbitals \cite{haw}. 
Because in our calculations we are concerned solely with states derived from the minimum of the CB, at $K,K'$ points, the tight-binding model used is sufficient.

We assume an armchair interface between materials -- see Fig.~\ref{fig:4}(a). Both termination types, armchair and zigzag are found to be stable in TMDC lateral heterostructures~\cite{oscar2}, however, zigzag termination, also analyzed by us, does not allow for proper valley-qubit transitions. This phenomenon will be analyzed in Section~\ref{sec:cond}.

\subsection{Poisson-Schr\"odinger method}

Voltages applied to the gates (relative to the substrate) are used to create the QD confinement potential in the monolayers area. Electric dipole charge at the junction is responsible for the built-in electric field. 
The confined electron itself also carries charge density (which we account for via the mean field approach).
To calculate the realistic electrostatic potential $\phi(\mathbf{r})$ we solve the generalized Poisson equation~\cite{mydrut}:
\begin{align}\label{eq:poiss}
    \boldsymbol{\nabla}\!\cdot\left(\varepsilon_0\varepsilon(\mathbf{r})\boldsymbol{\nabla}\Phi(\mathbf{r})\right)&=-\left(\rho_e(\mathbf{r})+d^{+}\!(\mathbf{r})+d^{-}\!(\mathbf{r})\right),\nonumber\\
    \phi(\mathbf{r})&=\Phi(\mathbf{r})-\phi_e(\mathbf{r}),
\end{align}
taking into account all of these components: voltages $V_\mathrm{\{L,R,U,D\}}$ applied to the control gates $\mathrm{G}_\mathrm{\{L,R,U,D\}}$ and to the highly doped substrate (kept at the referential potential $V_0=0$), space-dependent permittivity $\varepsilon(\mathbf{r})$ of different materials in the device (we assume $\varepsilon_\mathrm{MoS_2}=6.2$, $\varepsilon_\mathrm{WS_2}=6.1$, and $\varepsilon_\mathrm{hBN}=3.3$~\cite{diel}), electric dipole charge $d^\pm(\mathbf{r})$, and electron charge density $\rho_e(\mathbf{r})$ itself.
At the lateral and top sides of the computational box we apply Neumann boundary conditions with zeroing normal component of the electric field. 
To remove electron self-interaction we subtract electron potential itself $\phi_e(\mathbf{r})$ from the total potential $\Phi(\mathbf{r})$. Electron potential $\phi_e(\mathbf{r})$ is calculated using the standard Poisson equation: $\boldsymbol{\nabla}^2\phi_e(\mathbf{r})=-\rho_e(\mathbf{r})/(\varepsilon_0\varepsilon(\mathbf{r}))$ without any voltage conditions on the gates or the substrate.
Further details of the used method can be found in Ref.~[\onlinecite{mymos2}]. 

To calculate the electron eigenstates we solve the Schr\"odinger equation for the Hamiltonian (Eq.~\ref{ham1e}) with electrostatic potential $\phi(\mathbf{r})$ which is calculated via the Poisson equation,
\begin{equation}\label{eq:schr}
    H[\phi(\mathbf{r})]\psi_n(\mathbf{r})=E_n\psi_n(\mathbf{r}).
\end{equation}
On the other hand, to calculate the electrostatic potential one must solve the Poisson equation, which in turn requires knowledge of the electron charge density in a given QD state $\rho_e(\mathbf{r})=-|e||\psi_n(\mathbf{r})|^2$.
This means that both of the equations need to be solved self-consistently.

Presented in Fig.~\ref{fig:3} electron densities (dashed curves) for various QD-confinement configurations shows ground state density  $-|e||\psi_0(\mathbf{r})|^2$ calculated via the Poisson-Schr\"odinger system defined in Eqs.~\ref{eq:poiss} and \ref{eq:schr}.

During the time-dependent calculations we will be working in the calculated eigenstate basis, in which the full time-dependent wave function is represented as a linear combination of $N$ previously calculated basis states $\psi_n$:
\begin{equation}
\label{lin}
\Psi(\mathbf{r},t)=\sum_n c_n(t)\,\psi_n(\mathbf{r})e^{-\frac{\imath}{\hbar}E_n t},
\end{equation}
together with time-dependent amplitudes $c_n(t)$ and corresponding eigenvalues $E_n$. The time evolution is governed by the time-dependent Schr\"{o}dinger equation:
\begin{equation}
\label{schr}
\imath\hbar\frac{\partial}{\partial t} \Psi(\mathbf{r},t) = H(\mathbf{r},t) \Psi(\mathbf{r},t),
\end{equation}
with the time-dependent Hamiltonian being a sum of the stationary part (Eq.~\ref{ham1e}) and the variable contribution to the potential energy:
\begin{equation}
\label{var}
H(\mathbf{r},t)=H(\mathbf{r})-|e|\delta\phi(\mathbf{r},t).
\end{equation}
The full time-dependent potential $\phi(\mathbf{r},t)=\phi(\mathbf{r})+\delta\phi(\mathbf{r},t)$ contains variable part 
$\delta\phi(\mathbf{r},t)$, generated by modulation of the gate voltages. It is calculated by solving the Poisson equation for the variable density $\rho(\mathbf{r},t)$ at every time step. Note that the charge density originates from the actual wave-function, thus the Schr\"{o}dinger and Poisson equations are solved in a self-consistent way for each time step.  

Insertion of (\ref{lin}) to the Schr\"{o}dinger equation (\ref{schr}) gives a system of equations for time-derivatives of the expansion coefficients at subsequent moments in time:
\begin{equation}
\label{timeq}
\dot{c}_m(t)=-\frac{\imath}{\hbar}\sum_n c_n(t)\, \delta_{mn}(t)\, e^{\frac{\imath}{\hbar}(E_m-E_n)t}.
\end{equation}
The matrix elements $\delta_{mn}(t)=-|e|\langle \psi_m| \delta\phi(\mathbf{r},t)|\psi_n\rangle$ need to be calculated at every time step due to changes in the potential.

\section{Valley qubit manipulation}\label{sec:reslt}
After describing the computational method, let us apply voltages to the device gates $V_\mathrm{U}=V_\mathrm{D}=-2.5$~V, $V_\mathrm{L}=-1.2$~V, $V_\mathrm{R}=-2$~V and calculate self-consistently (via the Poisson-Schr\"odinger method) the QD confinement and electron ground state presented in Fig.~\ref{fig:4}(c,d)-left. Next, the time-dependent calculation starts. We change the left gate voltage to $V_\mathrm{L}=-1.5$~V, moving the QD confining potential (and electron density within) slightly to the flake center -- Fig.~\ref{fig:4}(c,d)-center. Then we start the pumping process by applying oscillatory voltages to the left gate: $V_\mathrm{L}(t)=-1.5-0.5(1\!-\!\cos(\omega t)/2$~V.
Oscillatory $V_\mathrm{L}(t)$ moves the electron back and forth towards the junction interface.
Respective potential profiles along the $x$-axis for different $V_\mathrm{L}$ voltages are also presented in Fig.~\ref{fig:3}(a).

\begin{figure}[t]
	\center
	\includegraphics[width=8.8cm]{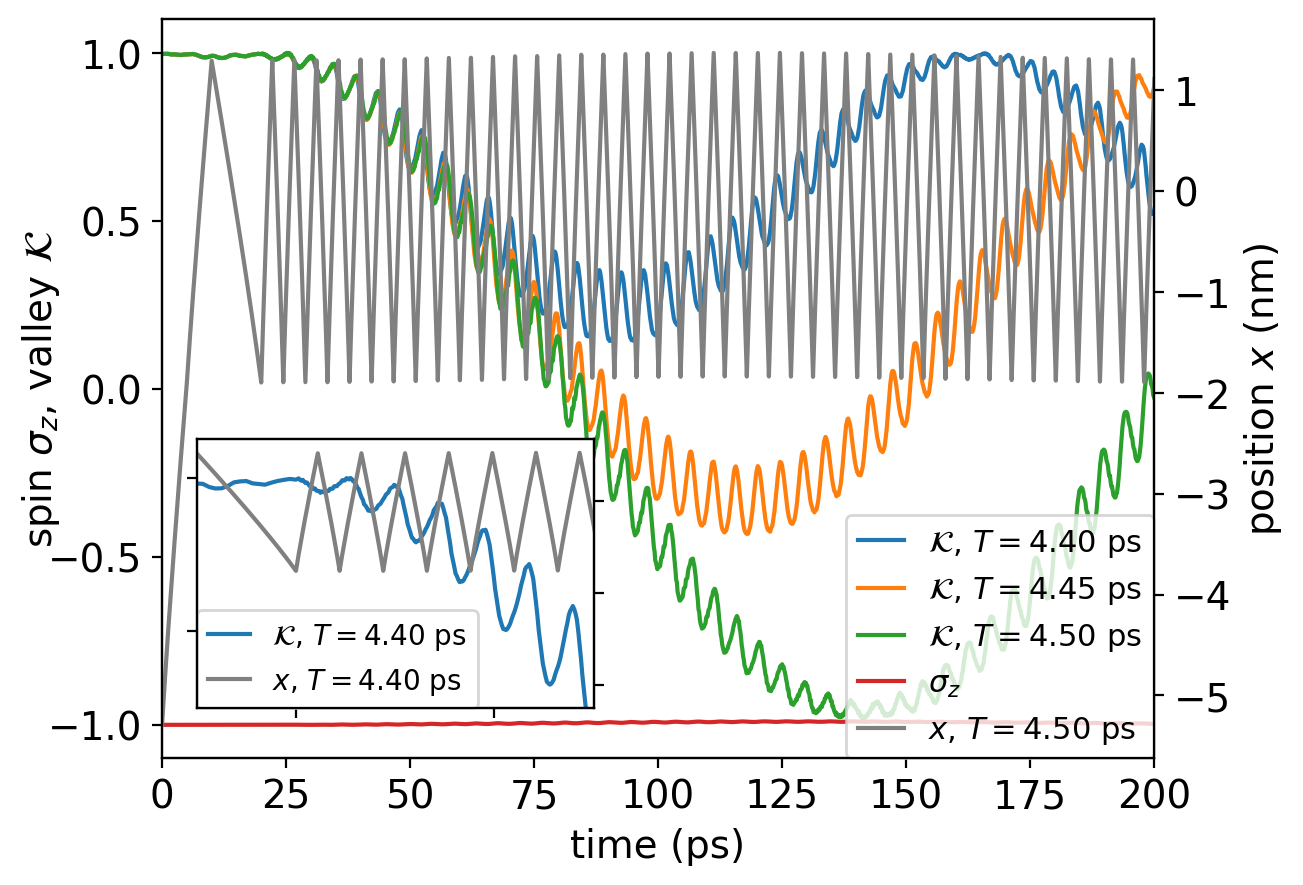}
	\caption{\label{fig:5} Dipole moment-induced intervalley transitions at the heterojunction interface for different frequencies $\omega$ of pumping that move the QD electron back and forth towards the junction. Transitions have a resonant nature, i.e., only a resonance period of $T=4.5$~ps gives full transitions.}
\end{figure}
When the pumping frequency $\omega$ is tuned to the qubit states' splitting (here, $2\pi/\omega=4.5$~ps), we start to observe the intervalley transitions presented in Fig.~\ref{fig:5}.
At $t=140$~ps, the valley isospin flips from $\mathcal{K}=-1$ ($K$~valley) to $\mathcal{K}=1$ ($K'$~valley).
These transitions are resonant and have the character of Rabi oscillations with incomplete transitions for out-of-resonance pumping frequency. 
The valley isospin value can be obtained by calculating the Fourier transform of the actual $\Psi(\mathbf{r},t)$ wavefunction, as presented in Fig.~\ref{fig:4}(f) -- details can be found in [\onlinecite{mymos2}].
Resonant pumping gradually transfers density from $K$ to $K'$ valley within the hexagonal BZ (white dashed line).

The frequency needed to address the qubit states' energy splitting of the order of meV reaches hundreds of GHz which might be problematic in any experimental setup. Luckily MoS$_2$ has relatively low spin-orbit CB energy splitting which combined with properly tuned perpendicular magnetic field may lead to much smaller splitting in given spin subspace (e.g. $\{|K\downarrow\rangle,|K'\downarrow\rangle\}$). This in turn can be addressed by much smaller pumping frequencies. Such situation occurs next to levels crossing about $B=2.4$~T shown in Fig.~\ref{fig:12}(b).
The low CB spin-orbit splitting in MoS$_2$ among all TMDC materials is the reason why it seems to be the most suitable for the proposed setup.

\begin{figure}[t]
	\center
\includegraphics[width=8.4cm]{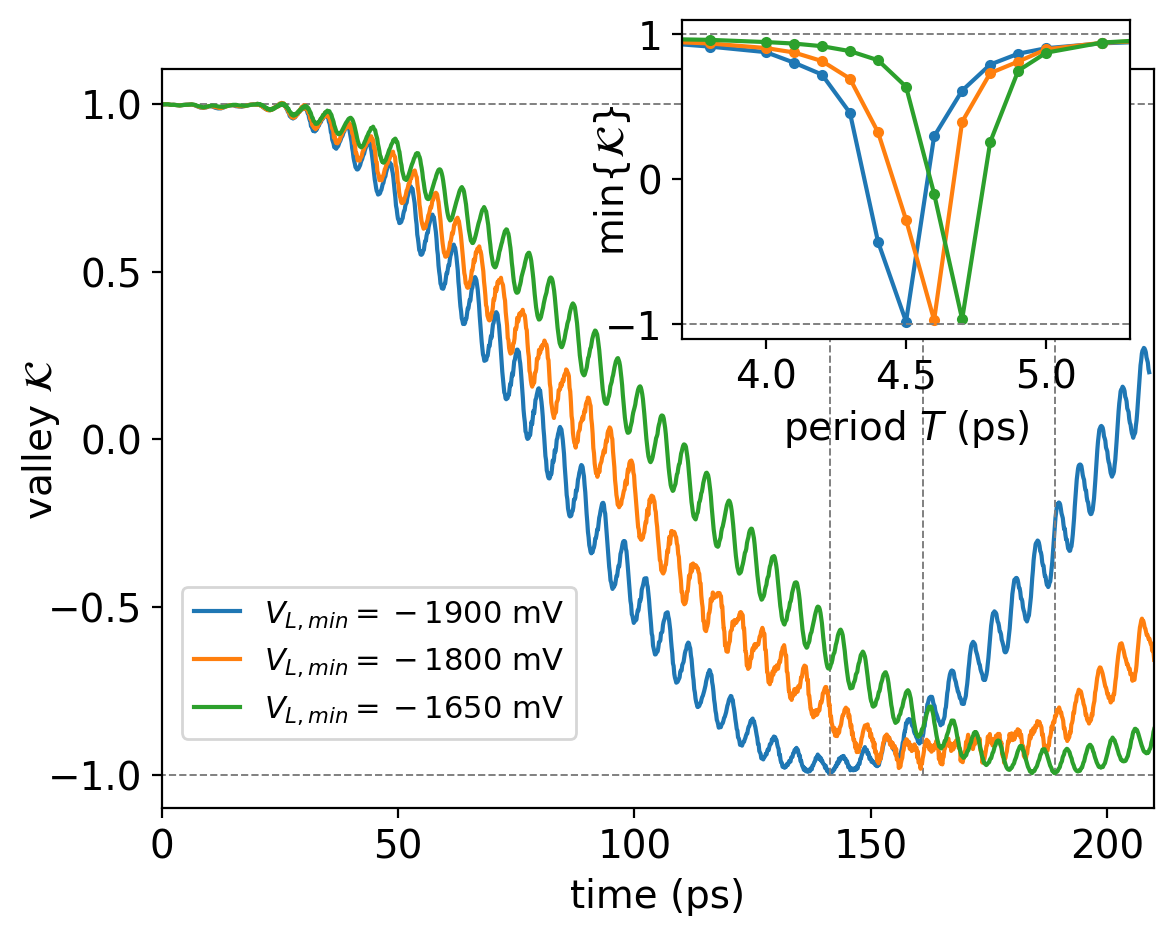}
	\caption{\label{fig:6} Intervalley transitions in the regime of weaker electron density overlapping with the junction interface, tuned by adjusting the $V_\mathrm{L,min}$ value. Inset: resonance curves for different pumping frequency/period.}
\end{figure}
To get transitions, the dipole moment alone is insufficient; the heterojunction interface must have proper termination. For a junction oriented with a zigzag interface we have not obtained transitions. All the results presented in Figs.~\ref{fig:5} and \ref{fig:6} were obtained for armchair termination between the MoS$_2$ and WS$_2$ monolayers.
This problem will be analyzed in Section~\ref{sec:cond}.

When the $V_\mathrm{L}(t)$ minimum value is higher (smaller $V_\mathrm{L}(t)$ oscillation amplitude) the junction's ability to induce transitions between valleys decreases. Presented in Fig~\ref{fig:6} are intervalley transitions with higher values of $V_\mathrm{L,min}>-2$~V, resulting in weaker overlapping with the junction interface and a
longer transition time (Rabi period) $> 140$~ps.
The resonant frequency is also slightly smaller (higher resonance period), and thus the resonance peak width is slightly reduced -- see inset in Fig~\ref{fig:6}.  

One should also note that in principle another definition of the qubit is possible~\cite{menaf}, i.e. using the two lowest spin-valley states. However, such configuration needs to have some spin-flip mechanism also present in the system, e.g. in a form of the Rashba coupling which is also included in our numerical model, in the Hamiltonian (Eq.~\ref{ham1e}).

\section{Intervalley transition conditions}\label{sec:cond}

There are two crucial conditions needed for intervalley transitions. The first one is related to the length scale of the linear dipole density at the junction interface, and the second one to the linear dipole orientation with respect to the monolayer lattice.

To meet the first condition, the perturbation, here in the form of the dipole potential, length scale should be comparable with the $K-K'$ difference in the momentum space~\cite{menaf,Sz_chenyi_2018}.
\begin{figure}[t]
	\center
	\includegraphics[width=8.4cm]{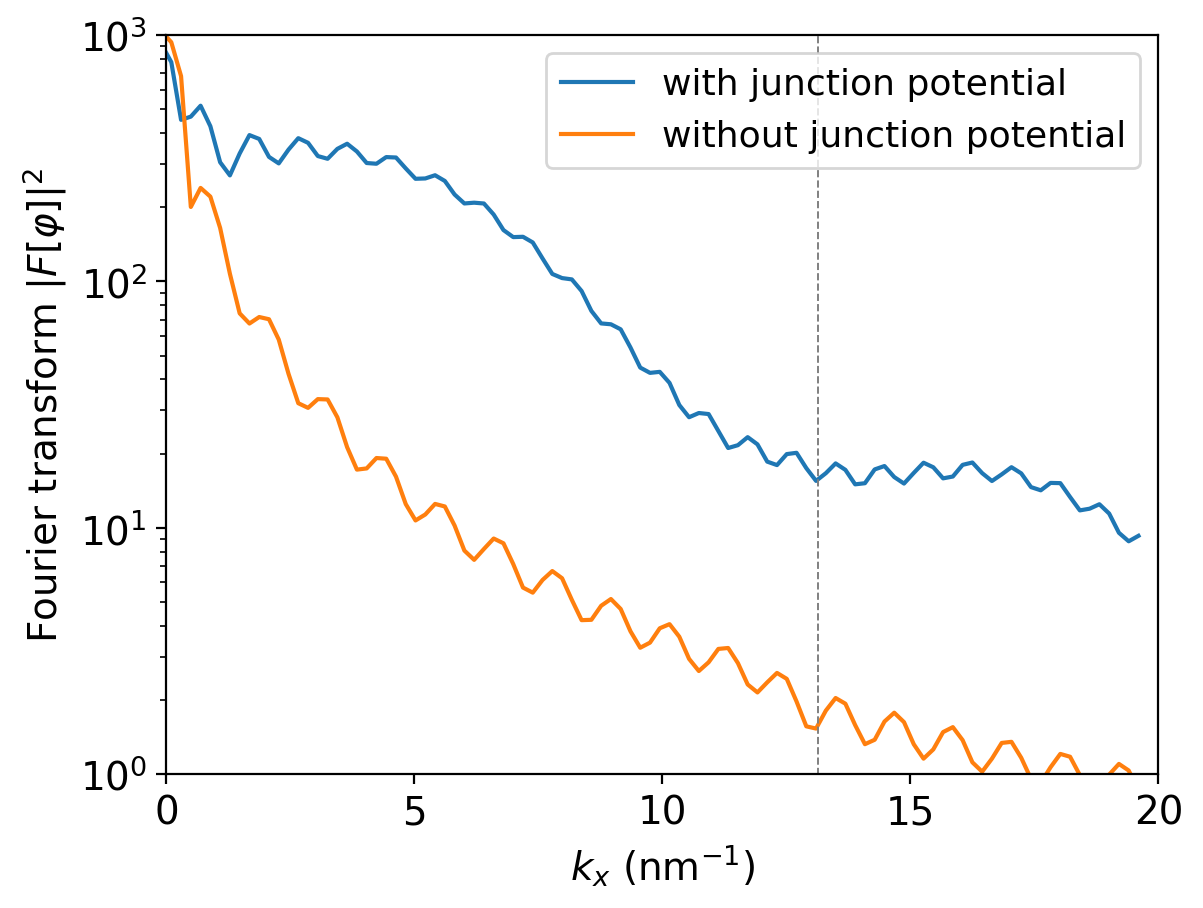}
	\caption{\label{fig:8} Fourier transform of the confinement potential with and without the junction potential (linear dipole).}
\end{figure}
This condition can be easily verified by calculating a Fourier transformation $F$ of the confinement $\varphi(x,y)=-|e|\phi(x,y)$. Presented in Fig.~\ref{fig:8} is the squared modulus of the 2D Fourier transform $|F[\varphi](k_x,k_y)|^2$ along the 
$k_y=0$ line 
($K'-K=(\frac{4\pi}{3a},0)$ -- see the BZ in Fig.~\ref{fig:4}(f), and the lattice constant $a=0.319$~nm)
for the original confinement potential (blue curve), and the same potential but with the junction (dipole) potential subtracted (orange). The latter one simply contains the gate-defined QD potential alone. It is clearly visible that amplitude at $k_x=\frac{4\pi}{3a}\simeq13.13$~nm$^{-1}$ is almost an order of magnitude larger for the case with a junction potential (more high-$k$-vector components due to rapidly varying area of the linear dipole) than in the case without a junction (only slowly varying QD potential).

The second condition states that the dipole orientation with respect to a monolayer lattice should be of armchair type.
\begin{figure}[b]
	\center
	\includegraphics[width=8.9cm]{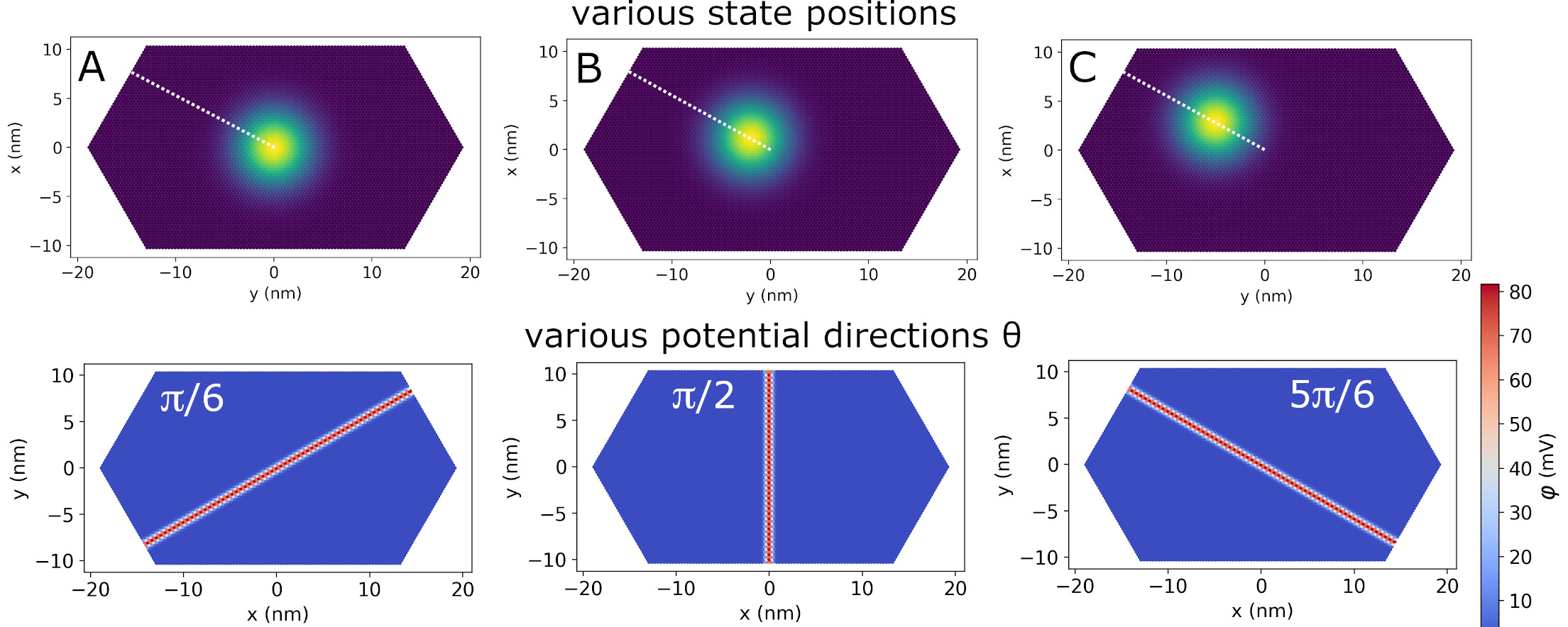}	\caption{\label{fig:9} (top) Various locations of the model Gaussian QD potential determine different positions of qubit state confined within. (bottom) Various directions of the Gaussian-shaped line potential that simulates the junction potential.}
\end{figure}
To study this problem we calculated potential matrix elements between the two basis states of the valley qubit: $\langle{}K|\varphi|K'\rangle$. 
To examine different junction orientations and qubit state positions with respect to the junction location, we assumed a model Gaussian-like potential and placed it in different positions, then we calculated eigenstates (using Eq.~\ref{ham1e}, but assuming MoS$_2$ material for the whole flake) and take two states $|K\rangle$ and $|K'\rangle$ from the CB minimum (with the same spin, e.g. oriented upwards) for three different locations (A,B,C) -- their densities are presented in Fig.~\ref{fig:9}(top).
Next, we define a Gaussian-shaped linear potential $\varphi$ that simulates the junction oriented along different directions $\theta$ -- as presented in Fig.~\ref{fig:9}(bottom).
The potential is positioned along a line passing through the flake center, but it can also be shifted laterally by some displacement $\delta$.
\begin{figure}[t]
	\center
	\includegraphics[width=8.7cm]{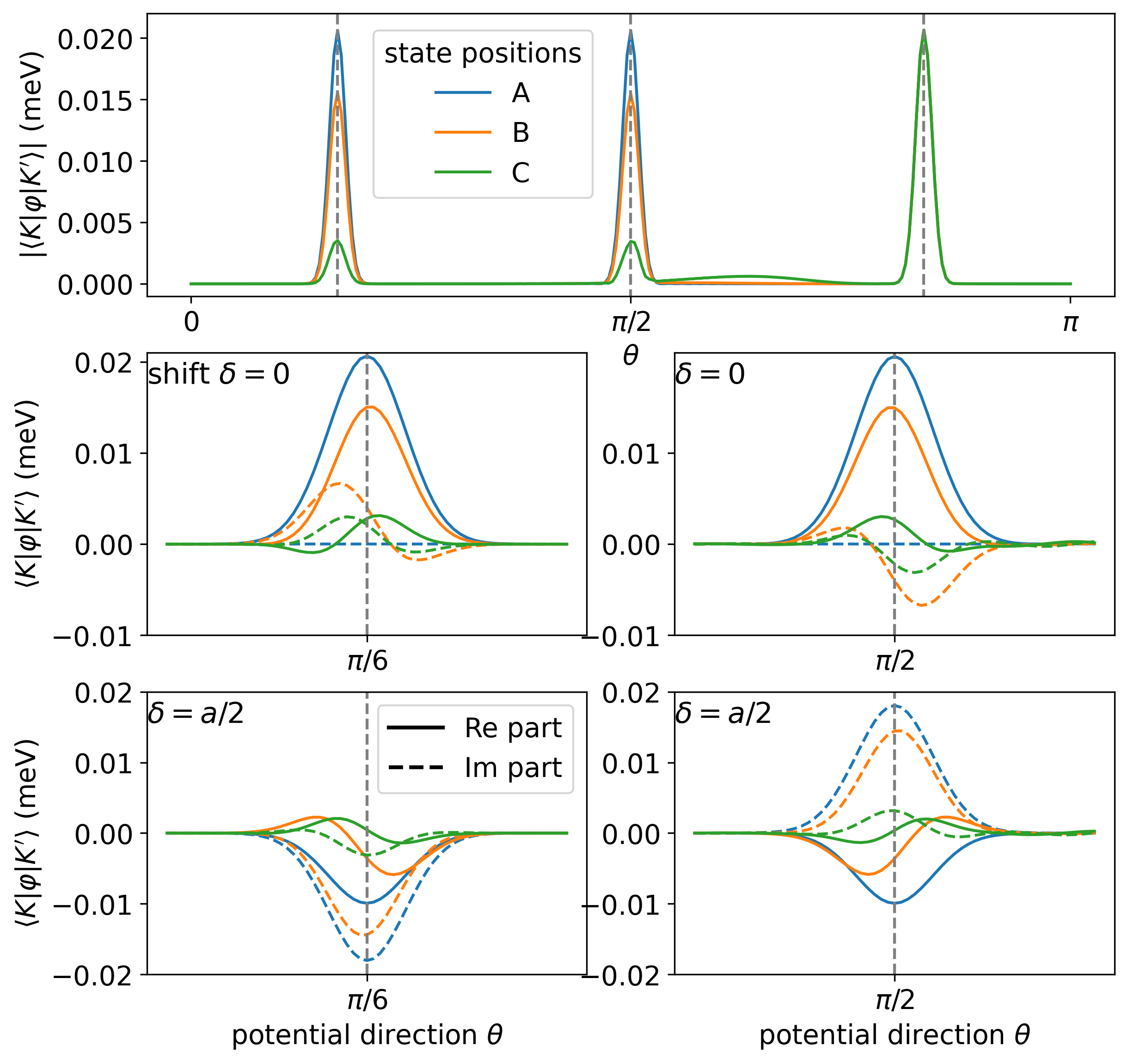}
	\caption{\label{fig:10} Matrix elements of artificially designed potential that mimics junction potential (oriented along different directions $\theta$ and with possible lateral shift $\delta$) for valley qubit states $|K\rangle$ and $|K'\rangle$ located at three different positions (A,B,C).}
\end{figure}
Finally, we calculated the matrix elements $\langle{}K|\varphi|K'\rangle$ for different junction orientations and qubit state positions.

Presented in Fig.~\ref{fig:10}(top) is the modulus of the matrix elements as a function of artificial junction orientation $\theta$ for various state positions. It is clearly visible that nonzero elements appear only next to $\theta=\frac{\pi}{6}, \frac{\pi}{2}, \frac{5\pi}{6}$, i.e. for a junction interface along the armchair termination. For example for $\theta=\frac{\pi}{2}$ we have the same termination as in Fig.~\ref{fig:4}(a). On the other hand, for zigzag terminations, i.e. $\theta=0, \frac{\pi}{3}, \frac{2\pi}{3}$, the matrix elements are zero for any state positions A-C. Also note that for $\theta=\frac{5\pi}{6}$, i.e. when states A-C are oriented along the junction, matrix elements are the same regardless of state's position.
Our findings that armchair termination may mix different valleys are in agreement with the present literature.
Armchair edges mix valleys and induce transitions in graphene-like structures~\cite{Brey2006,Zarenia2011,Szafran2018a,Szafran2018b}. Same edge-dependent valley mixing was proven for MoS$_2$ nanoribbons~\cite{Rostami2016}, or more generally for TMDC monolayers~\cite{Liu_2014}.

Armchair terminations give nonzero matrix elements for zero ($\delta=0$), or small nonzero ($\delta=a/2$) shifts as presented in Figs.~\ref{fig:10}(middle,bottom). It is also characteristic that detailed values of the matrix elements (real and imaginary part) strongly depend on the shift parameter, and for nonzero $\delta=a/2$ they have different form for different angles $\theta=\frac{\pi}{6}$ and $\theta=\frac{\pi}{2}$ -- especially imaginary parts are opposite -- see Fig.~\ref{fig:10}(bottom). For $\delta=0$ this dependence is a bit weaker but still visible, especially for the B and C state positions -- Fig.~\ref{fig:10}(middle).

\subsection{Analytical model for intervalley coupling}
To get better understanding of the underlying physics we also built a simple analytical model that captures only the features important to qualitatively describe the intervalley transitions mechanism -- sharp elongated potential aligned along a given direction with respect to the lattice vectors. Then we developed analytical formula for $\langle{}K|\varphi|K'\rangle$ element in that case. However, in order to estimate the detailed operation of the device, such as the optimal gate layout or the voltages needed to effectively control the qubit or its timings, it is necessary to get back to the original Poisson-tight-binding model.

To build the analytic model we assume 2D-Gaussian-like potential, centered at $\mathbf{x}_0=(x_0,y_0)^T$ and rotated by $\theta$, in a form:
\begin{align}\label{2dgauss}
\varphi_G(\mathbf{x}, \mathbf{x}_0,\theta,s_x,s_y)=U_0e^{-\frac{1}{2}(\mathbf{x}-\mathbf{x}_0)^T\!\mathbf{A}(\mathbf{x}-\mathbf{x}_0)},&\nonumber\\
\mathbf{A} = 
\begin{pmatrix}
    \frac{\cos^2\theta}{s^2_x}+\frac{\sin^2\theta}{s^2_y} & -\frac{\sin 2\theta}{s^2_x/2}+\frac{\sin 2\theta}{s^2_y/2} \\
    -\frac{\sin 2\theta}{s^2_x/2}+\frac{\sin 2\theta}{s^2_y/2} & \frac{\sin^2\theta}{s^2_x}+\frac{\cos^2\theta}{s^2_y} 
\end{pmatrix}&,
\end{align}
with $\mathbf{x}=(x,y)$ and standard deviations in both orthogonal directions defined as $(s_z,s_y)$.  

Now to calculate the Fourier transform of the $\varphi_G$ potential on a triangular lattice of Mo (or W) atoms, defined by lattice vectors $\mathbf{a}_1=a(1,0)$ and $\mathbf{a}_2=\frac{a}{2}(1,\sqrt{3})$, we have to perform the following summation over the whole lattice:
\begin{equation}\label{summ}
\varphi_G(\mathbf{q},\mathbf{k})=\frac{1}{N_{UC}}\sum_{j_1,j_2}e^{i(\mathbf{k}-\mathbf{q})(j_1\mathbf{a}_1+j_2\mathbf{a}_2)}\varphi_G(j_1\mathbf{a}_1+j_2\mathbf{a}_2),
\end{equation}
with $N_{UC}$ unit cells. For a finite lattice this sum is hard to be evaluated analytically, however, if one extend the system to infinite lattice the sum (\ref{summ}) can be approximated in continuum limit as ($\mathbf{Q}=\mathbf{k}-\mathbf{q}$):
\begin{align}\label{intt}
\varphi_G(\mathbf{Q})\simeq\int dj_1 dj_2e^{ia\left(Q_x(j_1 + j_2/2) + Q_y j_2\sqrt{3}/2\right)}&\times\nonumber\\
\times\varphi_G(a(j_1 + j_2/2), a j_2\sqrt{3}/2), x_0, y_0)&.
\end{align}
The integral (\ref{intt}) can be calculated using formula for generalized Gaussian integral (here for 2D case, $(j_1,j_2)$). After tedious calculations, and assuming that the potential $\varphi_G$ is centered at the position $(x_0,y_0)=(-\delta\sin\theta,\delta\cos\theta)$ where $\delta$ is the lateral displacement, one arrives at the formula:
\begin{align}
&\varphi_G(\mathbf{Q})\simeq\nonumber\\
&U_G e^{-\frac{1}{4}\left((s^2_x+s^2_y)(Q^2_x+Q^2_y)+(s^2_x-s^2_y)\left((Q^2_x-Q^2_y)\cos 2\theta+2Q_xQ_y \sin 2\theta\right)\right)}\nonumber\\
&\times e^{-i\delta\left(Q_x\sin\theta-Q_y\cos\theta\right)},
\end{align}
with $U_g=\frac{4\pi}{\sqrt{3}a^2}s_x s_y U_0$. If we lastly assume that the potential is significantly elongated in $x$ direction, i.e. $s_x\gg s_y$ (before rotation by $\theta$) we finally get:
\begin{align}
&\varphi_G(\mathbf{Q})\simeq U_G e^{-\frac{1}{4} \left(s^2_x(Q^2_x+Q^2_y)+s^2_x\left((Q^2_x-Q^2_y)\cos 2\theta+2Q_xQ_y \sin 2\theta\right)\right)}\nonumber\\
&\times e^{-i\delta\left(Q_x\sin\theta-Q_y\cos\theta\right)}.
\end{align}
The characteristic feature of $\varphi_G(\mathbf{Q})$ is that it is elongated, in $\mathbf{Q}$ space, in direction $\theta+\pi/2$, perpendicular to $\varphi_G(\mathbf{\mathbf{x}})$ in real space.
If we now calculate the matrix element $\langle K|\varphi_G|K'\rangle$ by putting $\mathbf{Q}=K'-K=(\frac{4\pi}{3a},0)$ we get a value which takes maximum at $\theta=-\frac{\pi}{2}$, i.e. the armchair direction:
\begin{equation}
\langle K|\varphi_G(\theta=-\pi/2)|K'\rangle=U_G e^{i\delta\frac{4\pi}{3a}}.
\end{equation}
Same values can be obtained for two other armchair directions, i.e. $\theta=\frac{\pi}{6}$ and $\frac{5\pi}{6}$:
\begin{equation}
\langle K|\varphi_G(\theta=\pi/6)|K'\rangle=\langle K|\varphi_G(\theta=5\pi/6)|K'\rangle=U_G e^{i\delta\frac{4\pi}{3a}}.
\end{equation}
These results are in agreement with numerical estimations from  Fig.~\ref{fig:10}.
If we then construct a junction to have a form similar to one from Fig.~\ref{fig:12}(a) we get:
\begin{align}
\langle K|\varphi_G(\theta=\pi/6)+\varphi_G(\theta=\pi/2)|K'\rangle
\sim&\nonumber\\
\sim\psi_\frac{\pi}{6}e^{i\delta\frac{4\pi}{3a}}+\psi_\frac{\pi}{2}e^{-i\delta\frac{4\pi}{3a}},
\end{align}
where $\psi_\frac{\pi}{6}$ is the amplitude for finding electron next to one arm (for $\theta=\pi/6$) of the junction, and $\psi_\frac{\pi}{2}$ for the other.
Now assume that the displacement $\delta$ is smaller than the lattice vector $a$ then the Hamiltonian in $\{K,K'\}$ basis takes the form:
\begin{equation}
H_{KK'}\sim\psi_\frac{\pi}{6}(\tau_x-K\delta\tau_y)+\psi_\frac{\pi}{2}(\tau_x+K\delta\tau_y),
\end{equation}
with $K=\frac{4\pi}{3a}$ and $\tau_i$ being valley Pauli matrices. What is obvious, operations connected to electron interaction with junction arms do not commute.

\section{Valley geometric phase}\label{sec:geom}
Results from Fig.~\ref{fig:10} motivated us to investigate the possibility of valley qubit manipulation via a geometric phase. 
Berry phase~\cite{Berry} naturally emerges in non-degenerate quantum state upon cyclic, adiabatic manipulation. It can also be generalized on degenerate systems~\cite{Wilczek} leading to so-called non-Abelian geometric phase. Such non-commutativity of spin rotation matrices, connected to adiabatic manipulation of electron spin, when it moves along a closed path, leads to an effective spin manipulation scheme~\cite{SanJose,Prabhakar,Pawlowski2012}. This scheme utilizes spin-orbit coupling, present in TMDC or III-V materials, and does not need any external magnetic field (degeneracy), which may limit qubit scalability. Resulting spin rotations also do not depend on dynamic evolution details, but only on the geometry of a given closed path.

When analysing results from Fig.~\ref{fig:10}, it became clear to us that intervalley matrix elements $\langle{}K|\varphi|K'\rangle$ for different junction angles (i.e., $\theta=\frac{\pi}{6}$ and $\theta=\frac{\pi}{2}$) may be interpreted as (compare middle panels and position C) $T_\frac{\pi}{6}=\frac{\eta}{2}(\tau_x-\tau_y)$ and $T_\frac{\pi}{2}=\frac{\eta}{2}(\tau_x+\tau_y)$ operators acting on the valley isospin subspace, with $\tau_i$ being valley Pauli matrices, and with some parameter $\eta$.
Since these two rotation generators do not commute, [$T_\frac{\pi}{6},T_\frac{\pi}{2}]=\eta^2 i\tau_z\neq 0$, we suppose that under proper arrangement of the junction, a nonzero geometric phase may emerge in our system, leading to effective valley manipulation.

\begin{figure}[t]
	\center
	\includegraphics[width=8.7cm]{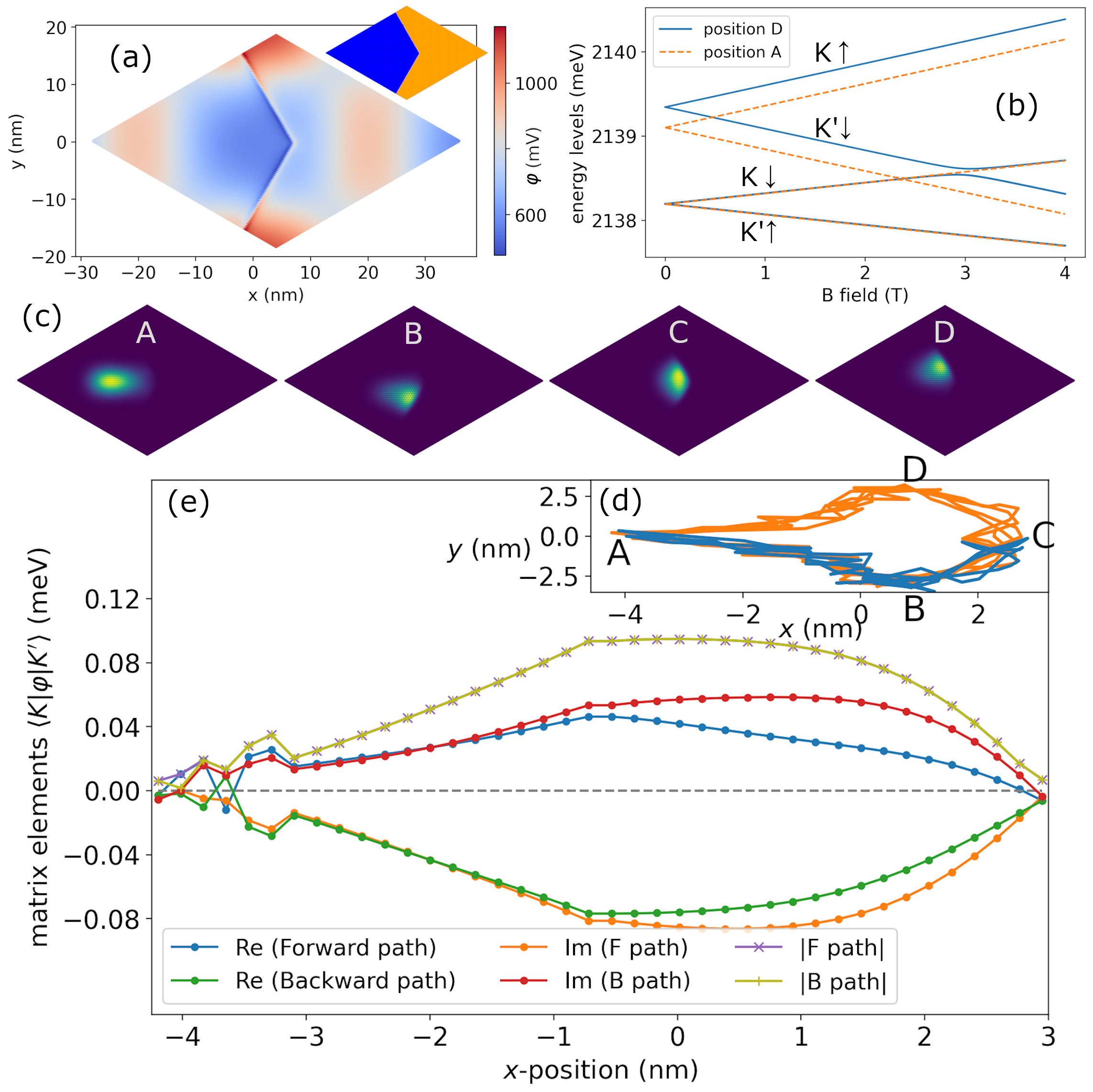}
	\caption{\label{fig:12} Setup designed for testing a possibility of valley qubit manipulation via geometric phase accumulation. It is based on corner-shaped junction (a) with both terminations of armchair type, while the phase accumulation occurs when circulating on a closed loop (d) near the junction.}
\end{figure}

To have the possibility non-commutative valley rotations and to make our analysis more realistic we planned another device presented in Fig.~\ref{fig:12}.
It contains a rhombus-shaped flake with crystal lattice oriented in such a way (rotated by $\frac{\pi}{6}$ with respect to the original lattice) that two armchair terminations can be arranged along $\theta=\frac{\pi}{3}$, and $\frac{2\pi}{3}$. Moreover, a lateral junction is formed along the crossing of such armchair terminations leading to a corner-shaped junction (with both sides of armchair type) as presented in Fig.~\ref{fig:12}(a). The corner junction is modelled in a similar manner as the original device with linear dipoles along the both armchair sides.

To enable us to work in a valley degenerate subspace, the external magnetic field $B=2.4$~T is applied to get appropriate level crossing (spin-down subspace: $|K\!\downarrow\rangle$ and $|K'\!\downarrow\rangle$ as presented by orange line in Fig.~\ref{fig:12}(b).
These levels were calculated for an electron localized in position A -- see Fig.~\ref{fig:12}(c). Moving the electron to other positions and therefore controlling its position is realized by applying control voltages to four device gates. This way we can adiabatically move the electron to positions B, C, and D, and after many cycles the electron path resembles a closed loop (orange curve) presented in Fig.~\ref{fig:12}(d).
In position A, the electron is farthest from the junction, while in positions B and D it is closest to the sides of the corner junction. When we calculate the energy levels as a function of the magnetic field for an electron in the D position, we may observe the influence of the nearby terminations manifested by the mixing (anticrossing) of energy levels -- blue line in Fig.~\ref{fig:12}(b).
Matrix elements calculated along the electron path presented in Fig.~\ref{fig:12}(e) show that forward (ABC path) and backward (CDA path) moves are represented by non commutative operators; e.g., at B position $T_\mathrm{forward}\simeq\eta'(\frac{1}{2}\sigma_x+\sigma_y)$ and $T_\mathrm{backward}\simeq-\eta'(\sigma_x+\frac{1}{2}\sigma_y)$.

\begin{figure}[t]
	\center
	\includegraphics[width=8.5cm]{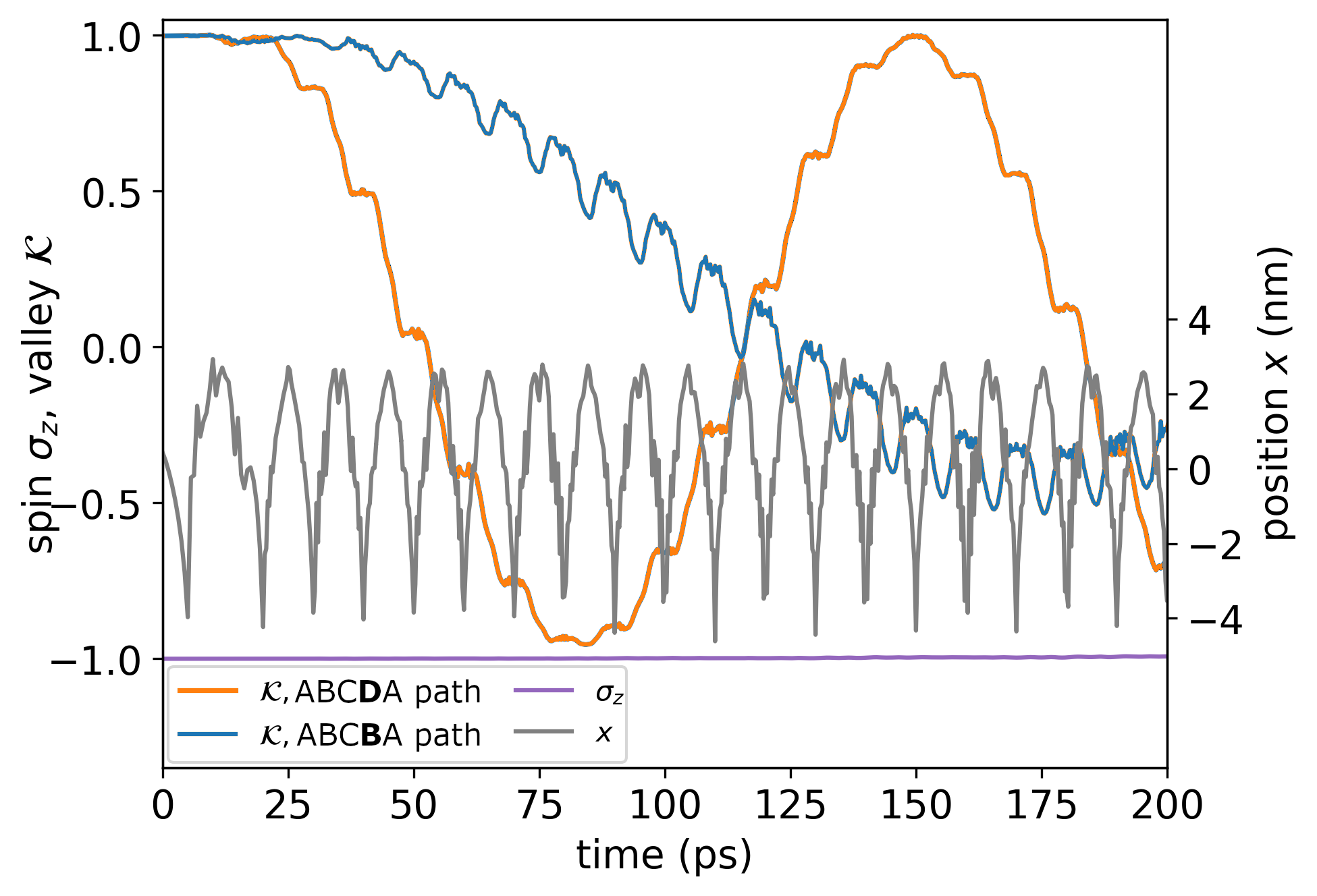}
	\caption{\label{fig:13} Valley isospin evolution when traveling along the closed path that enables accumulating of non-Abelian Berry phase (orange curve). When return path is the same, without circling the loop, the valley changes are much smaller (blue).}
\end{figure}
Now, let us verify if passing next to one side of a corner junction and then next to another induces effective valley isospin rotations; i.e., whether returning to initial position (ABCDA loop) results in the same or a rotated valley. Presented in Fig~\ref{fig:13} is the evolution of valley isospin of an electron confined in the nanodevice from Fig.~\ref{fig:12}. After passing ABCDA loop (blue curve) multiple times, one observes effective valley manipulation which gradually changes after the completion of each loop.
For comparison we also calculated valley evolution in the case of a return along the same path ABCBA (without making a loop -- see blue curve in Fig.~\ref{fig:12}(d)). In this case we still observe some valley changes (blue curve in Fig.~\ref{fig:13}) but with smaller amplitudes.

\section{Summary}
In the above work, we have studied a single-electron system in a TMDC lateral heterostructure, gate-defined quantum dot from the point of view of a valley-qubit implementation. Utilizing the time-dependent Schr\"{o}dinger equation in a tight-binding approximation coupled with the Poisson equation (which models a realistic dielectric environment), we were able to describe the proposed nanodevice with an in-plane TMDC heterojunction, four-gate QD geometry, and time-modulated electric potential.
We have shown that when the built-in dipole moment at the junction is oriented along the armchair interface, oscillatory pumping of the electron density into the junction area induces intervalley transitions. These transitions are resonant in nature and lead to Rabi oscillations with transition period dependent on electron strength overlapping with the junction area. 
Lateral interfaces between two different TMDC monolayers have been realized experimentally. However, the real interfaces are not always straight and frequently composed of shorter ($\sim10$~nm-long) sections of armchair and zigzag type~\cite{tmd-lateral4}. Thus, we also verified that intervalley coupling is possible even if the electron interacts only with a fragment (e.g. $\sim5$~nm-long) of armchair geometry (leaving the rest of such "kinked" junction in a zigzag form).
We also carefully analyzed what factors are necessary to achieve intervalley transitions. This analysis also led us to suggest another possibility of performing operations on the valley isospin. By properly designing a corner junction and carefully manipulating the electron to move along a closed loop near both sides of the junction, it is possible to accumulate the non-Abelian geometric phase and thus effectively rotate the valley.

Currently, another possibility to apply our scheme requiring sharp electronic potential modulations has emerged. Twisted TMDC bilayers seems to be a promising platform to study intervalley physics~\cite{twisted_tmd}. In moir\'{e} materials $K$ valleys are further splitted into two $K\pm$ points (forming so-called moir\'{e} BZ) that lie much closer to each other in $k$-space and are possibly easier to couple. Moreover, the moir\'{e} superlattice, that emerges in real space, may also play a role in inducing intervalley transitions. The benefit of this approach could lie in its compatibility with mechanically exfoliated TMDC crystals rather than epitaxially grown TMDCs, which are known to be of inferior quality.

\section{Acknowledgements}
The authors would like to thank Oscar \'Avalos-Ovando for enlightening discussions.
This work has been supported by National Science Centre, Poland, under Grant No. 2021/43/D/ST3/01989.
This research was supported in part by PL-Grid Infrastructure.

\bibliography{tmd-hetero}

\begin{thebibliography}{53}%
\makeatletter
\providecommand \@ifxundefined [1]{%
 \@ifx{#1\undefined}
}%
\providecommand \@ifnum [1]{%
 \ifnum #1\expandafter \@firstoftwo
 \else \expandafter \@secondoftwo
 \fi
}%
\providecommand \@ifx [1]{%
 \ifx #1\expandafter \@firstoftwo
 \else \expandafter \@secondoftwo
 \fi
}%
\providecommand \natexlab [1]{#1}%
\providecommand \enquote  [1]{``#1''}%
\providecommand \bibnamefont  [1]{#1}%
\providecommand \bibfnamefont [1]{#1}%
\providecommand \citenamefont [1]{#1}%
\providecommand \href@noop [0]{\@secondoftwo}%
\providecommand \href [0]{\begingroup \@sanitize@url \@href}%
\providecommand \@href[1]{\@@startlink{#1}\@@href}%
\providecommand \@@href[1]{\endgroup#1\@@endlink}%
\providecommand \@sanitize@url [0]{\catcode `\\12\catcode `\$12\catcode
  `\&12\catcode `\#12\catcode `\^12\catcode `\_12\catcode `\%12\relax}%
\providecommand \@@startlink[1]{}%
\providecommand \@@endlink[0]{}%
\providecommand \url  [0]{\begingroup\@sanitize@url \@url }%
\providecommand \@url [1]{\endgroup\@href {#1}{\urlprefix }}%
\providecommand \urlprefix  [0]{URL }%
\providecommand \Eprint [0]{\href }%
\providecommand \doibase [0]{https://doi.org/}%
\providecommand \selectlanguage [0]{\@gobble}%
\providecommand \bibinfo  [0]{\@secondoftwo}%
\providecommand \bibfield  [0]{\@secondoftwo}%
\providecommand \translation [1]{[#1]}%
\providecommand \BibitemOpen [0]{}%
\providecommand \bibitemStop [0]{}%
\providecommand \bibitemNoStop [0]{.\EOS\space}%
\providecommand \EOS [0]{\spacefactor3000\relax}%
\providecommand \BibitemShut  [1]{\csname bibitem#1\endcsname}%
\let\auto@bib@innerbib\@empty
\bibitem [{\citenamefont {Kjaergaard}\ \emph {et~al.}(2020)\citenamefont
  {Kjaergaard}, \citenamefont {Schwartz}, \citenamefont {Braum\"{u}ller},
  \citenamefont {Krantz}, \citenamefont {Wang}, \citenamefont {Gustavsson},\
  and\ \citenamefont {Oliver}}]{superconducting_current_state}%
  \BibitemOpen
  \bibfield  {author} {\bibinfo {author} {\bibfnamefont {M.}~\bibnamefont
  {Kjaergaard}}, \bibinfo {author} {\bibfnamefont {M.~E.}\ \bibnamefont
  {Schwartz}}, \bibinfo {author} {\bibfnamefont {J.}~\bibnamefont
  {Braum\"{u}ller}}, \bibinfo {author} {\bibfnamefont {P.}~\bibnamefont
  {Krantz}}, \bibinfo {author} {\bibfnamefont {J.~I.-J.}\ \bibnamefont {Wang}},
  \bibinfo {author} {\bibfnamefont {S.}~\bibnamefont {Gustavsson}},\ and\
  \bibinfo {author} {\bibfnamefont {W.~D.}\ \bibnamefont {Oliver}},\ }\href
  {https://doi.org/10.1146/annurev-conmatphys- 031119-050605} {\bibfield
  {journal} {\bibinfo  {journal} {Annual Review of Condensed Matter Physics}\
  }\textbf {\bibinfo {volume} {11}},\ \bibinfo {pages} {369} (\bibinfo {year}
  {2020})}\BibitemShut {NoStop}%
\bibitem [{\citenamefont {de~Leon}\ \emph {et~al.}(2021)\citenamefont
  {de~Leon}, \citenamefont {Itoh}, \citenamefont {Kim}, \citenamefont {Mehta},
  \citenamefont {Northup}, \citenamefont {Paik}, \citenamefont {Palmer},
  \citenamefont {Samarth}, \citenamefont {Sangtawesin},\ and\ \citenamefont
  {Steuerman}}]{science_material_challenges}%
  \BibitemOpen
  \bibfield  {author} {\bibinfo {author} {\bibfnamefont {N.~P.}\ \bibnamefont
  {de~Leon}}, \bibinfo {author} {\bibfnamefont {K.~M.}\ \bibnamefont {Itoh}},
  \bibinfo {author} {\bibfnamefont {D.}~\bibnamefont {Kim}}, \bibinfo {author}
  {\bibfnamefont {K.~K.}\ \bibnamefont {Mehta}}, \bibinfo {author}
  {\bibfnamefont {T.~E.}\ \bibnamefont {Northup}}, \bibinfo {author}
  {\bibfnamefont {H.}~\bibnamefont {Paik}}, \bibinfo {author} {\bibfnamefont
  {B.~S.}\ \bibnamefont {Palmer}}, \bibinfo {author} {\bibfnamefont
  {N.}~\bibnamefont {Samarth}}, \bibinfo {author} {\bibfnamefont
  {S.}~\bibnamefont {Sangtawesin}},\ and\ \bibinfo {author} {\bibfnamefont
  {D.~W.}\ \bibnamefont {Steuerman}},\ }\href
  {https://doi.org/10.1126/science.abb2823} {\bibfield  {journal} {\bibinfo
  {journal} {Science}\ }\textbf {\bibinfo {volume} {372}},\ \bibinfo {pages}
  {eabb2823} (\bibinfo {year} {2021})}\BibitemShut {NoStop}%
\bibitem [{\citenamefont {Watson}\ \emph {et~al.}(2018)\citenamefont {Watson},
  \citenamefont {Philips}, \citenamefont {Kawakami}, \citenamefont {Ward},
  \citenamefont {Scarlino}, \citenamefont {Veldhorst}, \citenamefont {Savage},
  \citenamefont {Lagally}, \citenamefont {Friesen}, \citenamefont
  {Coppersmith}, \citenamefont {Eriksson},\ and\ \citenamefont
  {Vandersypen}}]{programmable_two_qubit}%
  \BibitemOpen
  \bibfield  {author} {\bibinfo {author} {\bibfnamefont {T.~F.}\ \bibnamefont
  {Watson}}, \bibinfo {author} {\bibfnamefont {S.~G.~J.}\ \bibnamefont
  {Philips}}, \bibinfo {author} {\bibfnamefont {E.}~\bibnamefont {Kawakami}},
  \bibinfo {author} {\bibfnamefont {D.~R.}\ \bibnamefont {Ward}}, \bibinfo
  {author} {\bibfnamefont {P.}~\bibnamefont {Scarlino}}, \bibinfo {author}
  {\bibfnamefont {M.}~\bibnamefont {Veldhorst}}, \bibinfo {author}
  {\bibfnamefont {D.~E.}\ \bibnamefont {Savage}}, \bibinfo {author}
  {\bibfnamefont {M.~G.}\ \bibnamefont {Lagally}}, \bibinfo {author}
  {\bibfnamefont {M.}~\bibnamefont {Friesen}}, \bibinfo {author} {\bibfnamefont
  {S.~N.}\ \bibnamefont {Coppersmith}}, \bibinfo {author} {\bibfnamefont
  {M.~A.}\ \bibnamefont {Eriksson}},\ and\ \bibinfo {author} {\bibfnamefont
  {L.~M.~K.}\ \bibnamefont {Vandersypen}},\ }\href
  {https://doi.org/10.1038/nature25766} {\bibfield  {journal} {\bibinfo
  {journal} {Nature}\ }\textbf {\bibinfo {volume} {555}},\ \bibinfo {pages}
  {633} (\bibinfo {year} {2018})}\BibitemShut {NoStop}%
\bibitem [{\citenamefont {Lawrie}\ \emph {et~al.}(2020)\citenamefont {Lawrie},
  \citenamefont {Eenink}, \citenamefont {Hendrickx}, \citenamefont {Boter},
  \citenamefont {Petit}, \citenamefont {Amitonov}, \citenamefont {Lodari},
  \citenamefont {Paquelet~Wuetz}, \citenamefont {Volk}, \citenamefont
  {Philips}, \citenamefont {Droulers}, \citenamefont {Kalhor}, \citenamefont
  {van Riggelen}, \citenamefont {Brousse}, \citenamefont {Sammak},
  \citenamefont {Vandersypen}, \citenamefont {Scappucci},\ and\ \citenamefont
  {Veldhorst}}]{QD_array_Si_Ge}%
  \BibitemOpen
  \bibfield  {author} {\bibinfo {author} {\bibfnamefont {W.~I.~L.}\
  \bibnamefont {Lawrie}}, \bibinfo {author} {\bibfnamefont {H.~G.~J.}\
  \bibnamefont {Eenink}}, \bibinfo {author} {\bibfnamefont {N.~W.}\
  \bibnamefont {Hendrickx}}, \bibinfo {author} {\bibfnamefont {J.~M.}\
  \bibnamefont {Boter}}, \bibinfo {author} {\bibfnamefont {L.}~\bibnamefont
  {Petit}}, \bibinfo {author} {\bibfnamefont {S.~V.}\ \bibnamefont {Amitonov}},
  \bibinfo {author} {\bibfnamefont {M.}~\bibnamefont {Lodari}}, \bibinfo
  {author} {\bibfnamefont {B.}~\bibnamefont {Paquelet~Wuetz}}, \bibinfo
  {author} {\bibfnamefont {C.}~\bibnamefont {Volk}}, \bibinfo {author}
  {\bibfnamefont {S.~G.~J.}\ \bibnamefont {Philips}}, \bibinfo {author}
  {\bibfnamefont {G.}~\bibnamefont {Droulers}}, \bibinfo {author}
  {\bibfnamefont {N.}~\bibnamefont {Kalhor}}, \bibinfo {author} {\bibfnamefont
  {F.}~\bibnamefont {van Riggelen}}, \bibinfo {author} {\bibfnamefont
  {D.}~\bibnamefont {Brousse}}, \bibinfo {author} {\bibfnamefont
  {A.}~\bibnamefont {Sammak}}, \bibinfo {author} {\bibfnamefont {L.~M.~K.}\
  \bibnamefont {Vandersypen}}, \bibinfo {author} {\bibfnamefont
  {G.}~\bibnamefont {Scappucci}},\ and\ \bibinfo {author} {\bibfnamefont
  {M.}~\bibnamefont {Veldhorst}},\ }\href {https://doi.org/10.1063/5.0002013}
  {\bibfield  {journal} {\bibinfo  {journal} {Applied Physics Letters}\
  }\textbf {\bibinfo {volume} {116}},\ \bibinfo {pages} {080501} (\bibinfo
  {year} {2020})}\BibitemShut {NoStop}%
\bibitem [{\citenamefont {Penthorn}\ \emph {et~al.}(2019)\citenamefont
  {Penthorn}, \citenamefont {Schoenfield}, \citenamefont {Rooney},
  \citenamefont {Edge},\ and\ \citenamefont {Jiang}}]{two_axis_control}%
  \BibitemOpen
  \bibfield  {author} {\bibinfo {author} {\bibfnamefont {N.~E.}\ \bibnamefont
  {Penthorn}}, \bibinfo {author} {\bibfnamefont {J.~S.}\ \bibnamefont
  {Schoenfield}}, \bibinfo {author} {\bibfnamefont {J.~D.}\ \bibnamefont
  {Rooney}}, \bibinfo {author} {\bibfnamefont {L.~F.}\ \bibnamefont {Edge}},\
  and\ \bibinfo {author} {\bibfnamefont {H.}~\bibnamefont {Jiang}},\ }\href
  {https://doi.org/10.1038/s41534-019-0212-5} {\bibfield  {journal} {\bibinfo
  {journal} {npj Quantum Information}\ }\textbf {\bibinfo {volume} {5}},\
  \bibinfo {pages} {94} (\bibinfo {year} {2019})}\BibitemShut {NoStop}%
\bibitem [{\citenamefont {Li}\ \emph {et~al.}(2018)\citenamefont {Li},
  \citenamefont {Petit}, \citenamefont {Franke}, \citenamefont {Dehollain},
  \citenamefont {Helsen}, \citenamefont {Steudtner}, \citenamefont {Thomas},
  \citenamefont {Yoscovits}, \citenamefont {Singh}, \citenamefont {Wehner},
  \citenamefont {Vandersypen}, \citenamefont {Clarke},\ and\ \citenamefont
  {Veldhorst}}]{cross_bar_network}%
  \BibitemOpen
  \bibfield  {author} {\bibinfo {author} {\bibfnamefont {R.}~\bibnamefont
  {Li}}, \bibinfo {author} {\bibfnamefont {L.}~\bibnamefont {Petit}}, \bibinfo
  {author} {\bibfnamefont {D.~P.}\ \bibnamefont {Franke}}, \bibinfo {author}
  {\bibfnamefont {J.~P.}\ \bibnamefont {Dehollain}}, \bibinfo {author}
  {\bibfnamefont {J.}~\bibnamefont {Helsen}}, \bibinfo {author} {\bibfnamefont
  {M.}~\bibnamefont {Steudtner}}, \bibinfo {author} {\bibfnamefont {N.~K.}\
  \bibnamefont {Thomas}}, \bibinfo {author} {\bibfnamefont {Z.~R.}\
  \bibnamefont {Yoscovits}}, \bibinfo {author} {\bibfnamefont {K.~J.}\
  \bibnamefont {Singh}}, \bibinfo {author} {\bibfnamefont {S.}~\bibnamefont
  {Wehner}}, \bibinfo {author} {\bibfnamefont {L.~M.~K.}\ \bibnamefont
  {Vandersypen}}, \bibinfo {author} {\bibfnamefont {J.~S.}\ \bibnamefont
  {Clarke}},\ and\ \bibinfo {author} {\bibfnamefont {M.}~\bibnamefont
  {Veldhorst}},\ }\href {https://doi.org/10.1126/sciadv.aar3960} {\bibfield
  {journal} {\bibinfo  {journal} {Science Advances}\ }\textbf {\bibinfo
  {volume} {4}},\ \bibinfo {pages} {eaar3960} (\bibinfo {year}
  {2018})}\BibitemShut {NoStop}%
\bibitem [{\citenamefont {Lee}\ \emph {et~al.}(2020)\citenamefont {Lee},
  \citenamefont {Tsuchiya}, \citenamefont {Shinkai}, \citenamefont {Kanno},
  \citenamefont {Mine}, \citenamefont {Takahama}, \citenamefont {Mizokuchi},
  \citenamefont {Kodera}, \citenamefont {Hisamoto},\ and\ \citenamefont
  {Mizuno}}]{enhanced_electrostatic_coupling}%
  \BibitemOpen
  \bibfield  {author} {\bibinfo {author} {\bibfnamefont {N.}~\bibnamefont
  {Lee}}, \bibinfo {author} {\bibfnamefont {R.}~\bibnamefont {Tsuchiya}},
  \bibinfo {author} {\bibfnamefont {G.}~\bibnamefont {Shinkai}}, \bibinfo
  {author} {\bibfnamefont {Y.}~\bibnamefont {Kanno}}, \bibinfo {author}
  {\bibfnamefont {T.}~\bibnamefont {Mine}}, \bibinfo {author} {\bibfnamefont
  {T.}~\bibnamefont {Takahama}}, \bibinfo {author} {\bibfnamefont
  {R.}~\bibnamefont {Mizokuchi}}, \bibinfo {author} {\bibfnamefont
  {T.}~\bibnamefont {Kodera}}, \bibinfo {author} {\bibfnamefont
  {D.}~\bibnamefont {Hisamoto}},\ and\ \bibinfo {author} {\bibfnamefont
  {H.}~\bibnamefont {Mizuno}},\ }\href {https://doi.org/10.1063/1.5141522}
  {\bibfield  {journal} {\bibinfo  {journal} {Applied Physics Letters}\
  }\textbf {\bibinfo {volume} {116}},\ \bibinfo {pages} {162106} (\bibinfo
  {year} {2020})}\BibitemShut {NoStop}%
\bibitem [{\citenamefont {Korm\'anyos}\ \emph {et~al.}(2014)\citenamefont
  {Korm\'anyos}, \citenamefont {Z\'olyomi}, \citenamefont {Drummond},\ and\
  \citenamefont {Burkard}}]{Burkard2014}%
  \BibitemOpen
  \bibfield  {author} {\bibinfo {author} {\bibfnamefont {A.}~\bibnamefont
  {Korm\'anyos}}, \bibinfo {author} {\bibfnamefont {V.}~\bibnamefont
  {Z\'olyomi}}, \bibinfo {author} {\bibfnamefont {N.~D.}\ \bibnamefont
  {Drummond}},\ and\ \bibinfo {author} {\bibfnamefont {G.}~\bibnamefont
  {Burkard}},\ }\href {https://doi.org/10.1103/PhysRevX.4.011034} {\bibfield
  {journal} {\bibinfo  {journal} {Phys. Rev. X}\ }\textbf {\bibinfo {volume}
  {4}},\ \bibinfo {pages} {011034} (\bibinfo {year} {2014})}\BibitemShut
  {NoStop}%
\bibitem [{\citenamefont {Sierra}\ \emph {et~al.}(2021)\citenamefont {Sierra},
  \citenamefont {Fabian}, \citenamefont {Kawakami}, \citenamefont {Roche},\
  and\ \citenamefont {Valenzuela}}]{Fabian2021}%
  \BibitemOpen
  \bibfield  {author} {\bibinfo {author} {\bibfnamefont {J.~F.}\ \bibnamefont
  {Sierra}}, \bibinfo {author} {\bibfnamefont {J.}~\bibnamefont {Fabian}},
  \bibinfo {author} {\bibfnamefont {R.~K.}\ \bibnamefont {Kawakami}}, \bibinfo
  {author} {\bibfnamefont {S.}~\bibnamefont {Roche}},\ and\ \bibinfo {author}
  {\bibfnamefont {S.~O.}\ \bibnamefont {Valenzuela}},\ }\href
  {https://doi.org/10.1038/s41565-021-00936-x} {\bibfield  {journal} {\bibinfo
  {journal} {Nature Nanotechnology}\ }\textbf {\bibinfo {volume} {16}},\
  \bibinfo {pages} {856} (\bibinfo {year} {2021})}\BibitemShut {NoStop}%
\bibitem [{\citenamefont {Paw\l{}owski}\ \emph {et~al.}(2018)\citenamefont
  {Paw\l{}owski}, \citenamefont {\ifmmode~\dot{Z}\else \.{Z}\fi{}ebrowski},\
  and\ \citenamefont {Bednarek}}]{mymos2}%
  \BibitemOpen
  \bibfield  {author} {\bibinfo {author} {\bibfnamefont {J.}~\bibnamefont
  {Paw\l{}owski}}, \bibinfo {author} {\bibfnamefont {D.}~\bibnamefont
  {\ifmmode~\dot{Z}\else \.{Z}\fi{}ebrowski}},\ and\ \bibinfo {author}
  {\bibfnamefont {S.}~\bibnamefont {Bednarek}},\ }\href
  {https://doi.org/10.1103/PhysRevB.97.155412} {\bibfield  {journal} {\bibinfo
  {journal} {Phys. Rev. B}\ }\textbf {\bibinfo {volume} {97}},\ \bibinfo
  {pages} {155412} (\bibinfo {year} {2018})}\BibitemShut {NoStop}%
\bibitem [{\citenamefont {Paw{\l}owski}(2019)}]{myrash}%
  \BibitemOpen
  \bibfield  {author} {\bibinfo {author} {\bibfnamefont {J.}~\bibnamefont
  {Paw{\l}owski}},\ }\href {https://doi.org/10.1088/1367-2630/ab5ac9}
  {\bibfield  {journal} {\bibinfo  {journal} {New Journal of Physics}\ }\textbf
  {\bibinfo {volume} {21}},\ \bibinfo {pages} {123029} (\bibinfo {year}
  {2019})}\BibitemShut {NoStop}%
\bibitem [{\citenamefont {Brooks}\ and\ \citenamefont
  {Burkard}(2020)}]{Brooks2020}%
  \BibitemOpen
  \bibfield  {author} {\bibinfo {author} {\bibfnamefont {M.}~\bibnamefont
  {Brooks}}\ and\ \bibinfo {author} {\bibfnamefont {G.}~\bibnamefont
  {Burkard}},\ }\href {https://doi.org/10.1103/PhysRevB.101.035204} {\bibfield
  {journal} {\bibinfo  {journal} {Phys. Rev. B}\ }\textbf {\bibinfo {volume}
  {101}},\ \bibinfo {pages} {035204} (\bibinfo {year} {2020})}\BibitemShut
  {NoStop}%
\bibitem [{\citenamefont {Paw\l{}owski}\ \emph {et~al.}(2021)\citenamefont
  {Paw\l{}owski}, \citenamefont {Bieniek},\ and\ \citenamefont
  {Wo\ifmmode~\acute{z}\else \'{z}\fi{}niak}}]{Pawlowski2021}%
  \BibitemOpen
  \bibfield  {author} {\bibinfo {author} {\bibfnamefont {J.}~\bibnamefont
  {Paw\l{}owski}}, \bibinfo {author} {\bibfnamefont {M.}~\bibnamefont
  {Bieniek}},\ and\ \bibinfo {author} {\bibfnamefont {T.}~\bibnamefont
  {Wo\ifmmode~\acute{z}\else \'{z}\fi{}niak}},\ }\href
  {https://doi.org/10.1103/PhysRevApplied.15.054025} {\bibfield  {journal}
  {\bibinfo  {journal} {Phys. Rev. Applied}\ }\textbf {\bibinfo {volume}
  {15}},\ \bibinfo {pages} {054025} (\bibinfo {year} {2021})}\BibitemShut
  {NoStop}%
\bibitem [{\citenamefont {Alt\ifmmode \imath \else \i
  \fi{}nta\ifmmode~\mbox{\c{s}}\else \c{s}\fi{}}\ \emph
  {et~al.}(2021)\citenamefont {Alt\ifmmode \imath \else \i
  \fi{}nta\ifmmode~\mbox{\c{s}}\else \c{s}\fi{}}, \citenamefont {Bieniek},
  \citenamefont {Dusko}, \citenamefont {Korkusi\ifmmode~\acute{n}\else
  \'{n}\fi{}ski}, \citenamefont {Paw\l{}owski},\ and\ \citenamefont
  {Hawrylak}}]{menaf}%
  \BibitemOpen
  \bibfield  {author} {\bibinfo {author} {\bibfnamefont {A.}~\bibnamefont
  {Alt\ifmmode \imath \else \i \fi{}nta\ifmmode~\mbox{\c{s}}\else \c{s}\fi{}}},
  \bibinfo {author} {\bibfnamefont {M.}~\bibnamefont {Bieniek}}, \bibinfo
  {author} {\bibfnamefont {A.}~\bibnamefont {Dusko}}, \bibinfo {author}
  {\bibfnamefont {M.}~\bibnamefont {Korkusi\ifmmode~\acute{n}\else
  \'{n}\fi{}ski}}, \bibinfo {author} {\bibfnamefont {J.}~\bibnamefont
  {Paw\l{}owski}},\ and\ \bibinfo {author} {\bibfnamefont {P.}~\bibnamefont
  {Hawrylak}},\ }\href {https://doi.org/10.1103/PhysRevB.104.195412} {\bibfield
   {journal} {\bibinfo  {journal} {Phys. Rev. B}\ }\textbf {\bibinfo {volume}
  {104}},\ \bibinfo {pages} {195412} (\bibinfo {year} {2021})}\BibitemShut
  {NoStop}%
\bibitem [{\citenamefont {Boddison-Chouinard}\ \emph
  {et~al.}(2021)\citenamefont {Boddison-Chouinard}, \citenamefont {Bogan},
  \citenamefont {Fong}, \citenamefont {Watanabe}, \citenamefont {Taniguchi},
  \citenamefont {Studenikin}, \citenamefont {Sachrajda}, \citenamefont
  {Korkusinski}, \citenamefont {Altintas}, \citenamefont {Bieniek},
  \citenamefont {Hawrylak}, \citenamefont {Luican-Mayer},\ and\ \citenamefont
  {Gaudreau}}]{justin2021}%
  \BibitemOpen
  \bibfield  {author} {\bibinfo {author} {\bibfnamefont {J.}~\bibnamefont
  {Boddison-Chouinard}}, \bibinfo {author} {\bibfnamefont {A.}~\bibnamefont
  {Bogan}}, \bibinfo {author} {\bibfnamefont {N.}~\bibnamefont {Fong}},
  \bibinfo {author} {\bibfnamefont {K.}~\bibnamefont {Watanabe}}, \bibinfo
  {author} {\bibfnamefont {T.}~\bibnamefont {Taniguchi}}, \bibinfo {author}
  {\bibfnamefont {S.}~\bibnamefont {Studenikin}}, \bibinfo {author}
  {\bibfnamefont {A.}~\bibnamefont {Sachrajda}}, \bibinfo {author}
  {\bibfnamefont {M.}~\bibnamefont {Korkusinski}}, \bibinfo {author}
  {\bibfnamefont {A.}~\bibnamefont {Altintas}}, \bibinfo {author}
  {\bibfnamefont {M.}~\bibnamefont {Bieniek}}, \bibinfo {author} {\bibfnamefont
  {P.}~\bibnamefont {Hawrylak}}, \bibinfo {author} {\bibfnamefont
  {A.}~\bibnamefont {Luican-Mayer}},\ and\ \bibinfo {author} {\bibfnamefont
  {L.}~\bibnamefont {Gaudreau}},\ }\href {https://doi.org/10.1063/5.0062838}
  {\bibfield  {journal} {\bibinfo  {journal} {Applied Physics Letters}\
  }\textbf {\bibinfo {volume} {119}},\ \bibinfo {pages} {133104} (\bibinfo
  {year} {2021})}\BibitemShut {NoStop}%
\bibitem [{\citenamefont {{\'{A}}valos-Ovando}\ \emph
  {et~al.}(2019)\citenamefont {{\'{A}}valos-Ovando}, \citenamefont
  {Mastrogiuseppe},\ and\ \citenamefont {Ulloa}}]{oscar2}%
  \BibitemOpen
  \bibfield  {author} {\bibinfo {author} {\bibfnamefont {O.}~\bibnamefont
  {{\'{A}}valos-Ovando}}, \bibinfo {author} {\bibfnamefont {D.}~\bibnamefont
  {Mastrogiuseppe}},\ and\ \bibinfo {author} {\bibfnamefont {S.~E.}\
  \bibnamefont {Ulloa}},\ }\href {https://doi.org/10.1088/1361-648x/ab0970}
  {\bibfield  {journal} {\bibinfo  {journal} {Journal of Physics: Condensed
  Matter}\ }\textbf {\bibinfo {volume} {31}},\ \bibinfo {pages} {213001}
  (\bibinfo {year} {2019})}\BibitemShut {NoStop}%
\bibitem [{\citenamefont {Wang}\ \emph {et~al.}(2019)\citenamefont {Wang},
  \citenamefont {Li}, \citenamefont {Chen}, \citenamefont {Deng},\ and\
  \citenamefont {Niu}}]{recent_hetero}%
  \BibitemOpen
  \bibfield  {author} {\bibinfo {author} {\bibfnamefont {J.}~\bibnamefont
  {Wang}}, \bibinfo {author} {\bibfnamefont {Z.}~\bibnamefont {Li}}, \bibinfo
  {author} {\bibfnamefont {H.}~\bibnamefont {Chen}}, \bibinfo {author}
  {\bibfnamefont {G.}~\bibnamefont {Deng}},\ and\ \bibinfo {author}
  {\bibfnamefont {X.}~\bibnamefont {Niu}},\ }\href
  {https://doi.org/10.1007/s40820-019-0276-y} {\bibfield  {journal} {\bibinfo
  {journal} {Nano-Micro Letters}\ }\textbf {\bibinfo {volume} {11}},\ \bibinfo
  {pages} {48} (\bibinfo {year} {2019})}\BibitemShut {NoStop}%
\bibitem [{\citenamefont {Liu}\ \emph {et~al.}(2021)\citenamefont {Liu},
  \citenamefont {Wang}, \citenamefont {Liu}, \citenamefont {He}, \citenamefont
  {Chen}, \citenamefont {Li},\ and\ \citenamefont {Zhai}}]{small_structures}%
  \BibitemOpen
  \bibfield  {author} {\bibinfo {author} {\bibfnamefont {R.}~\bibnamefont
  {Liu}}, \bibinfo {author} {\bibfnamefont {F.}~\bibnamefont {Wang}}, \bibinfo
  {author} {\bibfnamefont {L.}~\bibnamefont {Liu}}, \bibinfo {author}
  {\bibfnamefont {X.}~\bibnamefont {He}}, \bibinfo {author} {\bibfnamefont
  {J.}~\bibnamefont {Chen}}, \bibinfo {author} {\bibfnamefont {Y.}~\bibnamefont
  {Li}},\ and\ \bibinfo {author} {\bibfnamefont {T.}~\bibnamefont {Zhai}},\
  }\href {https://doi.org/https://doi.org/10.1002/sstr.202000136} {\bibfield
  {journal} {\bibinfo  {journal} {Small Structures}\ }\textbf {\bibinfo
  {volume} {2}},\ \bibinfo {pages} {2000136} (\bibinfo {year}
  {2021})}\BibitemShut {NoStop}%
\bibitem [{\citenamefont {Bharadwaj}\ \emph
  {et~al.}(2022{\natexlab{a}})\citenamefont {Bharadwaj}, \citenamefont
  {Ramasubramaniam},\ and\ \citenamefont {Ram-Mohan}}]{satwik}%
  \BibitemOpen
  \bibfield  {author} {\bibinfo {author} {\bibfnamefont {S.}~\bibnamefont
  {Bharadwaj}}, \bibinfo {author} {\bibfnamefont {A.}~\bibnamefont
  {Ramasubramaniam}},\ and\ \bibinfo {author} {\bibfnamefont {L.~R.}\
  \bibnamefont {Ram-Mohan}},\ }\href {https://doi.org/10.1063/5.0089639}
  {\bibfield  {journal} {\bibinfo  {journal} {Journal of Applied Physics}\
  }\textbf {\bibinfo {volume} {131}},\ \bibinfo {pages} {174302} (\bibinfo
  {year} {2022}{\natexlab{a}})}\BibitemShut {NoStop}%
\bibitem [{\citenamefont {Duan}\ \emph {et~al.}(2014)\citenamefont {Duan},
  \citenamefont {Wang}, \citenamefont {Shaw}, \citenamefont {Cheng},
  \citenamefont {Chen}, \citenamefont {Li}, \citenamefont {Wu}, \citenamefont
  {Tang}, \citenamefont {Zhang}, \citenamefont {Pan}, \citenamefont {Jiang},
  \citenamefont {Yu}, \citenamefont {Huang},\ and\ \citenamefont
  {Duan}}]{tmd-lateral1}%
  \BibitemOpen
  \bibfield  {author} {\bibinfo {author} {\bibfnamefont {X.}~\bibnamefont
  {Duan}}, \bibinfo {author} {\bibfnamefont {C.}~\bibnamefont {Wang}}, \bibinfo
  {author} {\bibfnamefont {J.~C.}\ \bibnamefont {Shaw}}, \bibinfo {author}
  {\bibfnamefont {R.}~\bibnamefont {Cheng}}, \bibinfo {author} {\bibfnamefont
  {Y.}~\bibnamefont {Chen}}, \bibinfo {author} {\bibfnamefont {H.}~\bibnamefont
  {Li}}, \bibinfo {author} {\bibfnamefont {X.}~\bibnamefont {Wu}}, \bibinfo
  {author} {\bibfnamefont {Y.}~\bibnamefont {Tang}}, \bibinfo {author}
  {\bibfnamefont {Q.}~\bibnamefont {Zhang}}, \bibinfo {author} {\bibfnamefont
  {A.}~\bibnamefont {Pan}}, \bibinfo {author} {\bibfnamefont {J.}~\bibnamefont
  {Jiang}}, \bibinfo {author} {\bibfnamefont {R.}~\bibnamefont {Yu}}, \bibinfo
  {author} {\bibfnamefont {Y.}~\bibnamefont {Huang}},\ and\ \bibinfo {author}
  {\bibfnamefont {X.}~\bibnamefont {Duan}},\ }\href
  {https://doi.org/10.1038/nnano.2014.222} {\bibfield  {journal} {\bibinfo
  {journal} {Nature Nanotechnology}\ }\textbf {\bibinfo {volume} {9}},\
  \bibinfo {pages} {1024} (\bibinfo {year} {2014})}\BibitemShut {NoStop}%
\bibitem [{\citenamefont {Gong}\ \emph {et~al.}(2014)\citenamefont {Gong},
  \citenamefont {Lin}, \citenamefont {Wang}, \citenamefont {Shi}, \citenamefont
  {Lei}, \citenamefont {Lin}, \citenamefont {Zou}, \citenamefont {Ye},
  \citenamefont {Vajtai}, \citenamefont {Yakobson}, \citenamefont {Terrones},
  \citenamefont {Terrones}, \citenamefont {Tay}, \citenamefont {Lou},
  \citenamefont {Pantelides}, \citenamefont {Liu}, \citenamefont {Zhou},\ and\
  \citenamefont {Ajayan}}]{tmd-lateral2}%
  \BibitemOpen
  \bibfield  {author} {\bibinfo {author} {\bibfnamefont {Y.}~\bibnamefont
  {Gong}}, \bibinfo {author} {\bibfnamefont {J.}~\bibnamefont {Lin}}, \bibinfo
  {author} {\bibfnamefont {X.}~\bibnamefont {Wang}}, \bibinfo {author}
  {\bibfnamefont {G.}~\bibnamefont {Shi}}, \bibinfo {author} {\bibfnamefont
  {S.}~\bibnamefont {Lei}}, \bibinfo {author} {\bibfnamefont {Z.}~\bibnamefont
  {Lin}}, \bibinfo {author} {\bibfnamefont {X.}~\bibnamefont {Zou}}, \bibinfo
  {author} {\bibfnamefont {G.}~\bibnamefont {Ye}}, \bibinfo {author}
  {\bibfnamefont {R.}~\bibnamefont {Vajtai}}, \bibinfo {author} {\bibfnamefont
  {B.~I.}\ \bibnamefont {Yakobson}}, \bibinfo {author} {\bibfnamefont
  {H.}~\bibnamefont {Terrones}}, \bibinfo {author} {\bibfnamefont
  {M.}~\bibnamefont {Terrones}}, \bibinfo {author} {\bibfnamefont {B.~K.}\
  \bibnamefont {Tay}}, \bibinfo {author} {\bibfnamefont {J.}~\bibnamefont
  {Lou}}, \bibinfo {author} {\bibfnamefont {S.~T.}\ \bibnamefont {Pantelides}},
  \bibinfo {author} {\bibfnamefont {Z.}~\bibnamefont {Liu}}, \bibinfo {author}
  {\bibfnamefont {W.}~\bibnamefont {Zhou}},\ and\ \bibinfo {author}
  {\bibfnamefont {P.~M.}\ \bibnamefont {Ajayan}},\ }\href
  {https://doi.org/10.1038/nmat4091} {\bibfield  {journal} {\bibinfo  {journal}
  {Nature Materials}\ }\textbf {\bibinfo {volume} {13}},\ \bibinfo {pages}
  {1135} (\bibinfo {year} {2014})}\BibitemShut {NoStop}%
\bibitem [{\citenamefont {Huang}\ \emph {et~al.}(2014)\citenamefont {Huang},
  \citenamefont {Wu}, \citenamefont {Sanchez}, \citenamefont {Peters},
  \citenamefont {Beanland}, \citenamefont {Ross}, \citenamefont {Rivera},
  \citenamefont {Yao}, \citenamefont {Cobden},\ and\ \citenamefont
  {Xu}}]{tmd-lateral3}%
  \BibitemOpen
  \bibfield  {author} {\bibinfo {author} {\bibfnamefont {C.}~\bibnamefont
  {Huang}}, \bibinfo {author} {\bibfnamefont {S.}~\bibnamefont {Wu}}, \bibinfo
  {author} {\bibfnamefont {A.~M.}\ \bibnamefont {Sanchez}}, \bibinfo {author}
  {\bibfnamefont {J.~J.~P.}\ \bibnamefont {Peters}}, \bibinfo {author}
  {\bibfnamefont {R.}~\bibnamefont {Beanland}}, \bibinfo {author}
  {\bibfnamefont {J.~S.}\ \bibnamefont {Ross}}, \bibinfo {author}
  {\bibfnamefont {P.}~\bibnamefont {Rivera}}, \bibinfo {author} {\bibfnamefont
  {W.}~\bibnamefont {Yao}}, \bibinfo {author} {\bibfnamefont {D.~H.}\
  \bibnamefont {Cobden}},\ and\ \bibinfo {author} {\bibfnamefont
  {X.}~\bibnamefont {Xu}},\ }\href {https://doi.org/10.1038/nmat4064}
  {\bibfield  {journal} {\bibinfo  {journal} {Nature Materials}\ }\textbf
  {\bibinfo {volume} {13}},\ \bibinfo {pages} {1096} (\bibinfo {year}
  {2014})}\BibitemShut {NoStop}%
\bibitem [{\citenamefont {Li}\ \emph {et~al.}(2015{\natexlab{a}})\citenamefont
  {Li}, \citenamefont {Shi}, \citenamefont {Cheng}, \citenamefont {Lu},
  \citenamefont {Lin}, \citenamefont {Tang}, \citenamefont {Tsai},
  \citenamefont {Chu}, \citenamefont {Wei}, \citenamefont {He}, \citenamefont
  {Chang}, \citenamefont {Suenaga},\ and\ \citenamefont {Li}}]{tmd-lateral4}%
  \BibitemOpen
  \bibfield  {author} {\bibinfo {author} {\bibfnamefont {M.-Y.}\ \bibnamefont
  {Li}}, \bibinfo {author} {\bibfnamefont {Y.}~\bibnamefont {Shi}}, \bibinfo
  {author} {\bibfnamefont {C.-C.}\ \bibnamefont {Cheng}}, \bibinfo {author}
  {\bibfnamefont {L.-S.}\ \bibnamefont {Lu}}, \bibinfo {author} {\bibfnamefont
  {Y.-C.}\ \bibnamefont {Lin}}, \bibinfo {author} {\bibfnamefont {H.-L.}\
  \bibnamefont {Tang}}, \bibinfo {author} {\bibfnamefont {M.-L.}\ \bibnamefont
  {Tsai}}, \bibinfo {author} {\bibfnamefont {C.-W.}\ \bibnamefont {Chu}},
  \bibinfo {author} {\bibfnamefont {K.-H.}\ \bibnamefont {Wei}}, \bibinfo
  {author} {\bibfnamefont {J.-H.}\ \bibnamefont {He}}, \bibinfo {author}
  {\bibfnamefont {W.-H.}\ \bibnamefont {Chang}}, \bibinfo {author}
  {\bibfnamefont {K.}~\bibnamefont {Suenaga}},\ and\ \bibinfo {author}
  {\bibfnamefont {L.-J.}\ \bibnamefont {Li}},\ }\href
  {https://doi.org/10.1126/science.aab4097} {\bibfield  {journal} {\bibinfo
  {journal} {Science}\ }\textbf {\bibinfo {volume} {349}},\ \bibinfo {pages}
  {524} (\bibinfo {year} {2015}{\natexlab{a}})}\BibitemShut {NoStop}%
\bibitem [{\citenamefont {Zhang}\ \emph {et~al.}(2015)\citenamefont {Zhang},
  \citenamefont {Lin}, \citenamefont {Tseng}, \citenamefont {Huang},\ and\
  \citenamefont {Lee}}]{tmd-lateral5}%
  \BibitemOpen
  \bibfield  {author} {\bibinfo {author} {\bibfnamefont {X.-Q.}\ \bibnamefont
  {Zhang}}, \bibinfo {author} {\bibfnamefont {C.-H.}\ \bibnamefont {Lin}},
  \bibinfo {author} {\bibfnamefont {Y.-W.}\ \bibnamefont {Tseng}}, \bibinfo
  {author} {\bibfnamefont {K.-H.}\ \bibnamefont {Huang}},\ and\ \bibinfo
  {author} {\bibfnamefont {Y.-H.}\ \bibnamefont {Lee}},\ }\href
  {https://doi.org/10.1021/nl503744f} {\bibfield  {journal} {\bibinfo
  {journal} {Nano Letters}\ }\textbf {\bibinfo {volume} {15}},\ \bibinfo
  {pages} {410} (\bibinfo {year} {2015})}\BibitemShut {NoStop}%
\bibitem [{\citenamefont {Zhang}\ \emph {et~al.}(2018)\citenamefont {Zhang},
  \citenamefont {Li}, \citenamefont {Tersoff}, \citenamefont {Han},
  \citenamefont {Su}, \citenamefont {Li}, \citenamefont {Muller},\ and\
  \citenamefont {Shih}}]{wse2_strain}%
  \BibitemOpen
  \bibfield  {author} {\bibinfo {author} {\bibfnamefont {C.}~\bibnamefont
  {Zhang}}, \bibinfo {author} {\bibfnamefont {M.-Y.}\ \bibnamefont {Li}},
  \bibinfo {author} {\bibfnamefont {J.}~\bibnamefont {Tersoff}}, \bibinfo
  {author} {\bibfnamefont {Y.}~\bibnamefont {Han}}, \bibinfo {author}
  {\bibfnamefont {Y.}~\bibnamefont {Su}}, \bibinfo {author} {\bibfnamefont
  {L.-J.}\ \bibnamefont {Li}}, \bibinfo {author} {\bibfnamefont {D.~A.}\
  \bibnamefont {Muller}},\ and\ \bibinfo {author} {\bibfnamefont {C.-K.}\
  \bibnamefont {Shih}},\ }\href {https://doi.org/10.1038/s41565-017-0022-x}
  {\bibfield  {journal} {\bibinfo  {journal} {Nature Nanotechnology}\ }\textbf
  {\bibinfo {volume} {13}},\ \bibinfo {pages} {152} (\bibinfo {year}
  {2018})}\BibitemShut {NoStop}%
\bibitem [{\citenamefont {Bharadwaj}\ \emph
  {et~al.}(2022{\natexlab{b}})\citenamefont {Bharadwaj}, \citenamefont
  {Ramasubramaniam},\ and\ \citenamefont {Ram-Mohan}}]{thermo}%
  \BibitemOpen
  \bibfield  {author} {\bibinfo {author} {\bibfnamefont {S.}~\bibnamefont
  {Bharadwaj}}, \bibinfo {author} {\bibfnamefont {A.}~\bibnamefont
  {Ramasubramaniam}},\ and\ \bibinfo {author} {\bibfnamefont {L.~R.}\
  \bibnamefont {Ram-Mohan}},\ }\href {https://doi.org/10.1039/D2NR01609E}
  {\bibfield  {journal} {\bibinfo  {journal} {Nanoscale}\ }\textbf {\bibinfo
  {volume} {14}},\ \bibinfo {pages} {11750} (\bibinfo {year}
  {2022}{\natexlab{b}})}\BibitemShut {NoStop}%
\bibitem [{\citenamefont {Wu}\ \emph {et~al.}(2016)\citenamefont {Wu},
  \citenamefont {Tong}, \citenamefont {Liu}, \citenamefont {Yu},\ and\
  \citenamefont {Yao}}]{Yue2016}%
  \BibitemOpen
  \bibfield  {author} {\bibinfo {author} {\bibfnamefont {Y.}~\bibnamefont
  {Wu}}, \bibinfo {author} {\bibfnamefont {Q.}~\bibnamefont {Tong}}, \bibinfo
  {author} {\bibfnamefont {G.-B.}\ \bibnamefont {Liu}}, \bibinfo {author}
  {\bibfnamefont {H.}~\bibnamefont {Yu}},\ and\ \bibinfo {author}
  {\bibfnamefont {W.}~\bibnamefont {Yao}},\ }\href
  {https://doi.org/10.1103/PhysRevB.93.045313} {\bibfield  {journal} {\bibinfo
  {journal} {Phys. Rev. B}\ }\textbf {\bibinfo {volume} {93}},\ \bibinfo
  {pages} {045313} (\bibinfo {year} {2016})}\BibitemShut {NoStop}%
\bibitem [{\citenamefont {Liu}\ \emph {et~al.}(2014)\citenamefont {Liu},
  \citenamefont {Pang}, \citenamefont {Yao},\ and\ \citenamefont
  {Yao}}]{Liu_2014}%
  \BibitemOpen
  \bibfield  {author} {\bibinfo {author} {\bibfnamefont {G.-B.}\ \bibnamefont
  {Liu}}, \bibinfo {author} {\bibfnamefont {H.}~\bibnamefont {Pang}}, \bibinfo
  {author} {\bibfnamefont {Y.}~\bibnamefont {Yao}},\ and\ \bibinfo {author}
  {\bibfnamefont {W.}~\bibnamefont {Yao}},\ }\href
  {https://doi.org/10.1088/1367-2630/16/10/105011} {\bibfield  {journal}
  {\bibinfo  {journal} {New Journal of Physics}\ }\textbf {\bibinfo {volume}
  {16}},\ \bibinfo {pages} {105011} (\bibinfo {year} {2014})}\BibitemShut
  {NoStop}%
\bibitem [{\citenamefont {Sz{\'{e}}chenyi}\ \emph {et~al.}(2018)\citenamefont
  {Sz{\'{e}}chenyi}, \citenamefont {Chirolli},\ and\ \citenamefont
  {P{\'{a}}lyi}}]{Sz_chenyi_2018}%
  \BibitemOpen
  \bibfield  {author} {\bibinfo {author} {\bibfnamefont {G.}~\bibnamefont
  {Sz{\'{e}}chenyi}}, \bibinfo {author} {\bibfnamefont {L.}~\bibnamefont
  {Chirolli}},\ and\ \bibinfo {author} {\bibfnamefont {A.}~\bibnamefont
  {P{\'{a}}lyi}},\ }\href {https://doi.org/10.1088/2053-1583/aab80e} {\bibfield
   {journal} {\bibinfo  {journal} {2D Materials}\ }\textbf {\bibinfo {volume}
  {5}},\ \bibinfo {pages} {035004} (\bibinfo {year} {2018})}\BibitemShut
  {NoStop}%
\bibitem [{\citenamefont {Lee}\ \emph {et~al.}(2017)\citenamefont {Lee},
  \citenamefont {Huang}, \citenamefont {Sumpter},\ and\ \citenamefont
  {Yoon}}]{Lee_2017}%
  \BibitemOpen
  \bibfield  {author} {\bibinfo {author} {\bibfnamefont {J.}~\bibnamefont
  {Lee}}, \bibinfo {author} {\bibfnamefont {J.}~\bibnamefont {Huang}}, \bibinfo
  {author} {\bibfnamefont {B.~G.}\ \bibnamefont {Sumpter}},\ and\ \bibinfo
  {author} {\bibfnamefont {M.}~\bibnamefont {Yoon}},\ }\href
  {https://doi.org/10.1088/2053-1583/aa5542} {\bibfield  {journal} {\bibinfo
  {journal} {2D Materials}\ }\textbf {\bibinfo {volume} {4}},\ \bibinfo {pages}
  {021016} (\bibinfo {year} {2017})}\BibitemShut {NoStop}%
\bibitem [{\citenamefont {Kang}\ \emph {et~al.}(2013)\citenamefont {Kang},
  \citenamefont {Tongay}, \citenamefont {Zhou}, \citenamefont {Li},\ and\
  \citenamefont {Wu}}]{junction1}%
  \BibitemOpen
  \bibfield  {author} {\bibinfo {author} {\bibfnamefont {J.}~\bibnamefont
  {Kang}}, \bibinfo {author} {\bibfnamefont {S.}~\bibnamefont {Tongay}},
  \bibinfo {author} {\bibfnamefont {J.}~\bibnamefont {Zhou}}, \bibinfo {author}
  {\bibfnamefont {J.}~\bibnamefont {Li}},\ and\ \bibinfo {author}
  {\bibfnamefont {J.}~\bibnamefont {Wu}},\ }\href
  {https://doi.org/10.1063/1.4774090} {\bibfield  {journal} {\bibinfo
  {journal} {Applied Physics Letters}\ }\textbf {\bibinfo {volume} {102}},\
  \bibinfo {pages} {012111} (\bibinfo {year} {2013})}\BibitemShut {NoStop}%
\bibitem [{\citenamefont {Wei}\ \emph {et~al.}(2016)\citenamefont {Wei},
  \citenamefont {Dai},\ and\ \citenamefont {Huang}}]{junction2}%
  \BibitemOpen
  \bibfield  {author} {\bibinfo {author} {\bibfnamefont {W.}~\bibnamefont
  {Wei}}, \bibinfo {author} {\bibfnamefont {Y.}~\bibnamefont {Dai}},\ and\
  \bibinfo {author} {\bibfnamefont {B.}~\bibnamefont {Huang}},\ }\href
  {https://doi.org/10.1039/C6CP02741E} {\bibfield  {journal} {\bibinfo
  {journal} {Phys. Chem. Chem. Phys.}\ }\textbf {\bibinfo {volume} {18}},\
  \bibinfo {pages} {15632} (\bibinfo {year} {2016})}\BibitemShut {NoStop}%
\bibitem [{\citenamefont {Wen}\ \emph {et~al.}(2019)\citenamefont {Wen},
  \citenamefont {He}, \citenamefont {Wang}, \citenamefont {Yao}, \citenamefont
  {Zhang}, \citenamefont {Hussain}, \citenamefont {Wang}, \citenamefont
  {Cheng}, \citenamefont {Yin}, \citenamefont {Getaye~Sendeku}, \citenamefont
  {Wang}, \citenamefont {Jiang},\ and\ \citenamefont {He}}]{cht1}%
  \BibitemOpen
  \bibfield  {author} {\bibinfo {author} {\bibfnamefont {Y.}~\bibnamefont
  {Wen}}, \bibinfo {author} {\bibfnamefont {P.}~\bibnamefont {He}}, \bibinfo
  {author} {\bibfnamefont {Q.}~\bibnamefont {Wang}}, \bibinfo {author}
  {\bibfnamefont {Y.}~\bibnamefont {Yao}}, \bibinfo {author} {\bibfnamefont
  {Y.}~\bibnamefont {Zhang}}, \bibinfo {author} {\bibfnamefont
  {S.}~\bibnamefont {Hussain}}, \bibinfo {author} {\bibfnamefont
  {Z.}~\bibnamefont {Wang}}, \bibinfo {author} {\bibfnamefont {R.}~\bibnamefont
  {Cheng}}, \bibinfo {author} {\bibfnamefont {L.}~\bibnamefont {Yin}}, \bibinfo
  {author} {\bibfnamefont {M.}~\bibnamefont {Getaye~Sendeku}}, \bibinfo
  {author} {\bibfnamefont {F.}~\bibnamefont {Wang}}, \bibinfo {author}
  {\bibfnamefont {C.}~\bibnamefont {Jiang}},\ and\ \bibinfo {author}
  {\bibfnamefont {J.}~\bibnamefont {He}},\ }\bibfield  {booktitle} {\emph
  {\bibinfo {booktitle} {ACS Nano}},\ }\href
  {https://doi.org/10.1021/acsnano.9b08375} {\bibfield  {journal} {\bibinfo
  {journal} {ACS Nano}\ }\textbf {\bibinfo {volume} {13}},\ \bibinfo {pages}
  {14519} (\bibinfo {year} {2019})}\BibitemShut {NoStop}%
\bibitem [{\citenamefont {Li}\ \emph {et~al.}(2015{\natexlab{b}})\citenamefont
  {Li}, \citenamefont {Yu}, \citenamefont {Li}, \citenamefont {Tang},
  \citenamefont {Guo}, \citenamefont {Lei}, \citenamefont {Fu},\ and\
  \citenamefont {Zhu}}]{cht2}%
  \BibitemOpen
  \bibfield  {author} {\bibinfo {author} {\bibfnamefont {H.}~\bibnamefont
  {Li}}, \bibinfo {author} {\bibfnamefont {K.}~\bibnamefont {Yu}}, \bibinfo
  {author} {\bibfnamefont {C.}~\bibnamefont {Li}}, \bibinfo {author}
  {\bibfnamefont {Z.}~\bibnamefont {Tang}}, \bibinfo {author} {\bibfnamefont
  {B.}~\bibnamefont {Guo}}, \bibinfo {author} {\bibfnamefont {X.}~\bibnamefont
  {Lei}}, \bibinfo {author} {\bibfnamefont {H.}~\bibnamefont {Fu}},\ and\
  \bibinfo {author} {\bibfnamefont {Z.}~\bibnamefont {Zhu}},\ }\href
  {https://doi.org/10.1038/srep18730} {\bibfield  {journal} {\bibinfo
  {journal} {Scientific Reports}\ }\textbf {\bibinfo {volume} {5}},\ \bibinfo
  {pages} {18730} (\bibinfo {year} {2015}{\natexlab{b}})}\BibitemShut {NoStop}%
\bibitem [{\citenamefont {Zhang}\ \emph {et~al.}(2016)\citenamefont {Zhang},
  \citenamefont {Xie}, \citenamefont {Zhao},\ and\ \citenamefont
  {Zhang}}]{junction3}%
  \BibitemOpen
  \bibfield  {author} {\bibinfo {author} {\bibfnamefont {J.}~\bibnamefont
  {Zhang}}, \bibinfo {author} {\bibfnamefont {W.}~\bibnamefont {Xie}}, \bibinfo
  {author} {\bibfnamefont {J.}~\bibnamefont {Zhao}},\ and\ \bibinfo {author}
  {\bibfnamefont {S.}~\bibnamefont {Zhang}},\ }\href
  {https://doi.org/10.1088/2053-1583/aa50cc} {\bibfield  {journal} {\bibinfo
  {journal} {2D Materials}\ }\textbf {\bibinfo {volume} {4}},\ \bibinfo {pages}
  {015038} (\bibinfo {year} {2016})}\BibitemShut {NoStop}%
\bibitem [{\citenamefont {\'Avalos-Ovando}\ \emph {et~al.}(2019)\citenamefont
  {\'Avalos-Ovando}, \citenamefont {Mastrogiuseppe},\ and\ \citenamefont
  {Ulloa}}]{oscar1}%
  \BibitemOpen
  \bibfield  {author} {\bibinfo {author} {\bibfnamefont {O.}~\bibnamefont
  {\'Avalos-Ovando}}, \bibinfo {author} {\bibfnamefont {D.}~\bibnamefont
  {Mastrogiuseppe}},\ and\ \bibinfo {author} {\bibfnamefont {S.~E.}\
  \bibnamefont {Ulloa}},\ }\href {https://doi.org/10.1103/PhysRevB.99.035107}
  {\bibfield  {journal} {\bibinfo  {journal} {Phys. Rev. B}\ }\textbf {\bibinfo
  {volume} {99}},\ \bibinfo {pages} {035107} (\bibinfo {year}
  {2019})}\BibitemShut {NoStop}%
\bibitem [{\citenamefont {Ko\ifmmode~\acute{s}\else \'{s}\fi{}mider}\ \emph
  {et~al.}(2013)\citenamefont {Ko\ifmmode~\acute{s}\else \'{s}\fi{}mider},
  \citenamefont {Gonz\'alez},\ and\ \citenamefont {Fern\'andez-Rossier}}]{kos}%
  \BibitemOpen
  \bibfield  {author} {\bibinfo {author} {\bibfnamefont {K.}~\bibnamefont
  {Ko\ifmmode~\acute{s}\else \'{s}\fi{}mider}}, \bibinfo {author}
  {\bibfnamefont {J.~W.}\ \bibnamefont {Gonz\'alez}},\ and\ \bibinfo {author}
  {\bibfnamefont {J.}~\bibnamefont {Fern\'andez-Rossier}},\ }\href
  {https://doi.org/10.1103/PhysRevB.88.245436} {\bibfield  {journal} {\bibinfo
  {journal} {Phys. Rev. B}\ }\textbf {\bibinfo {volume} {88}},\ \bibinfo
  {pages} {245436} (\bibinfo {year} {2013})}\BibitemShut {NoStop}%
\bibitem [{\citenamefont {Choukroun}\ \emph {et~al.}(2018)\citenamefont
  {Choukroun}, \citenamefont {Pala}, \citenamefont {Fang}, \citenamefont
  {Kaxiras},\ and\ \citenamefont {Dollfus}}]{half}%
  \BibitemOpen
  \bibfield  {author} {\bibinfo {author} {\bibfnamefont {J.}~\bibnamefont
  {Choukroun}}, \bibinfo {author} {\bibfnamefont {M.}~\bibnamefont {Pala}},
  \bibinfo {author} {\bibfnamefont {S.}~\bibnamefont {Fang}}, \bibinfo {author}
  {\bibfnamefont {E.}~\bibnamefont {Kaxiras}},\ and\ \bibinfo {author}
  {\bibfnamefont {P.}~\bibnamefont {Dollfus}},\ }\href
  {https://doi.org/10.1088/1361-6528/aae7df} {\bibfield  {journal} {\bibinfo
  {journal} {Nanotechnology}\ }\textbf {\bibinfo {volume} {30}},\ \bibinfo
  {pages} {025201} (\bibinfo {year} {2018})}\BibitemShut {NoStop}%
\bibitem [{\citenamefont {Liu}\ \emph {et~al.}(2013)\citenamefont {Liu},
  \citenamefont {Shan}, \citenamefont {Yao}, \citenamefont {Yao},\ and\
  \citenamefont {Xiao}}]{xiao}%
  \BibitemOpen
  \bibfield  {author} {\bibinfo {author} {\bibfnamefont {G.-B.}\ \bibnamefont
  {Liu}}, \bibinfo {author} {\bibfnamefont {W.-Y.}\ \bibnamefont {Shan}},
  \bibinfo {author} {\bibfnamefont {Y.}~\bibnamefont {Yao}}, \bibinfo {author}
  {\bibfnamefont {W.}~\bibnamefont {Yao}},\ and\ \bibinfo {author}
  {\bibfnamefont {D.}~\bibnamefont {Xiao}},\ }\href
  {https://doi.org/10.1103/PhysRevB.88.085433} {\bibfield  {journal} {\bibinfo
  {journal} {Phys. Rev. B}\ }\textbf {\bibinfo {volume} {88}},\ \bibinfo
  {pages} {085433} (\bibinfo {year} {2013})}\BibitemShut {NoStop}%
\bibitem [{\citenamefont {Kadantsev}\ and\ \citenamefont
  {Hawrylak}(2012)}]{haw}%
  \BibitemOpen
  \bibfield  {author} {\bibinfo {author} {\bibfnamefont {E.~S.}\ \bibnamefont
  {Kadantsev}}\ and\ \bibinfo {author} {\bibfnamefont {P.}~\bibnamefont
  {Hawrylak}},\ }\href {https://doi.org/10.1016/j.ssc.2012.02.005} {\bibfield
  {journal} {\bibinfo  {journal} {Solid State Communications}\ }\textbf
  {\bibinfo {volume} {152}},\ \bibinfo {pages} {909} (\bibinfo {year}
  {2012})}\BibitemShut {NoStop}%
\bibitem [{\citenamefont {Paw\l{}owski}\ \emph {et~al.}(2016)\citenamefont
  {Paw\l{}owski}, \citenamefont {Szumniak},\ and\ \citenamefont
  {Bednarek}}]{mydrut}%
  \BibitemOpen
  \bibfield  {author} {\bibinfo {author} {\bibfnamefont {J.}~\bibnamefont
  {Paw\l{}owski}}, \bibinfo {author} {\bibfnamefont {P.}~\bibnamefont
  {Szumniak}},\ and\ \bibinfo {author} {\bibfnamefont {S.}~\bibnamefont
  {Bednarek}},\ }\href {https://doi.org/10.1103/PhysRevB.93.045309} {\bibfield
  {journal} {\bibinfo  {journal} {Phys. Rev. B}\ }\textbf {\bibinfo {volume}
  {93}},\ \bibinfo {pages} {045309} (\bibinfo {year} {2016})}\BibitemShut
  {NoStop}%
\bibitem [{\citenamefont {Laturia}\ \emph {et~al.}(2018)\citenamefont
  {Laturia}, \citenamefont {Van~de Put},\ and\ \citenamefont
  {Vandenberghe}}]{diel}%
  \BibitemOpen
  \bibfield  {author} {\bibinfo {author} {\bibfnamefont {A.}~\bibnamefont
  {Laturia}}, \bibinfo {author} {\bibfnamefont {M.~L.}\ \bibnamefont {Van~de
  Put}},\ and\ \bibinfo {author} {\bibfnamefont {W.~G.}\ \bibnamefont
  {Vandenberghe}},\ }\href@noop {} {\bibfield  {journal} {\bibinfo  {journal}
  {npj 2D Materials and Applications}\ }\textbf {\bibinfo {volume} {2}},\
  \bibinfo {pages} {6} (\bibinfo {year} {2018})}\BibitemShut {NoStop}%
\bibitem [{\citenamefont {Brey}\ and\ \citenamefont {Fertig}(2006)}]{Brey2006}%
  \BibitemOpen
  \bibfield  {author} {\bibinfo {author} {\bibfnamefont {L.}~\bibnamefont
  {Brey}}\ and\ \bibinfo {author} {\bibfnamefont {H.~A.}\ \bibnamefont
  {Fertig}},\ }\href {https://doi.org/10.1103/PhysRevB.73.235411} {\bibfield
  {journal} {\bibinfo  {journal} {Phys. Rev. B}\ }\textbf {\bibinfo {volume}
  {73}},\ \bibinfo {pages} {235411} (\bibinfo {year} {2006})}\BibitemShut
  {NoStop}%
\bibitem [{\citenamefont {Zarenia}\ \emph {et~al.}(2011)\citenamefont
  {Zarenia}, \citenamefont {Chaves}, \citenamefont {Farias},\ and\
  \citenamefont {Peeters}}]{Zarenia2011}%
  \BibitemOpen
  \bibfield  {author} {\bibinfo {author} {\bibfnamefont {M.}~\bibnamefont
  {Zarenia}}, \bibinfo {author} {\bibfnamefont {A.}~\bibnamefont {Chaves}},
  \bibinfo {author} {\bibfnamefont {G.~A.}\ \bibnamefont {Farias}},\ and\
  \bibinfo {author} {\bibfnamefont {F.~M.}\ \bibnamefont {Peeters}},\ }\href
  {https://doi.org/10.1103/PhysRevB.84.245403} {\bibfield  {journal} {\bibinfo
  {journal} {Phys. Rev. B}\ }\textbf {\bibinfo {volume} {84}},\ \bibinfo
  {pages} {245403} (\bibinfo {year} {2011})}\BibitemShut {NoStop}%
\bibitem [{\citenamefont {Szafran}\ \emph
  {et~al.}(2018{\natexlab{a}})\citenamefont {Szafran}, \citenamefont
  {{\.Z}ebrowski},\ and\ \citenamefont
  {Mre{\'n}ca-Kolasi{\'n}ska}}]{Szafran2018a}%
  \BibitemOpen
  \bibfield  {author} {\bibinfo {author} {\bibfnamefont {B.}~\bibnamefont
  {Szafran}}, \bibinfo {author} {\bibfnamefont {D.}~\bibnamefont
  {{\.Z}ebrowski}},\ and\ \bibinfo {author} {\bibfnamefont {A.}~\bibnamefont
  {Mre{\'n}ca-Kolasi{\'n}ska}},\ }\href
  {https://doi.org/10.1038/s41598-018-25534-1} {\bibfield  {journal} {\bibinfo
  {journal} {Scientific Reports}\ }\textbf {\bibinfo {volume} {8}},\ \bibinfo
  {pages} {7166} (\bibinfo {year} {2018}{\natexlab{a}})}\BibitemShut {NoStop}%
\bibitem [{\citenamefont {Szafran}\ \emph
  {et~al.}(2018{\natexlab{b}})\citenamefont {Szafran}, \citenamefont
  {Mre\ifmmode \acute{n}\else \'{n}\fi{}ca-Kolasi\ifmmode~\acute{n}\else
  \'{n}\fi{}ska}, \citenamefont {Rzeszotarski},\ and\ \citenamefont
  {\ifmmode~\dot{Z}\else \.{Z}\fi{}ebrowski}}]{Szafran2018b}%
  \BibitemOpen
  \bibfield  {author} {\bibinfo {author} {\bibfnamefont {B.}~\bibnamefont
  {Szafran}}, \bibinfo {author} {\bibfnamefont {A.}~\bibnamefont {Mre\ifmmode
  \acute{n}\else \'{n}\fi{}ca-Kolasi\ifmmode~\acute{n}\else \'{n}\fi{}ska}},
  \bibinfo {author} {\bibfnamefont {B.}~\bibnamefont {Rzeszotarski}},\ and\
  \bibinfo {author} {\bibfnamefont {D.}~\bibnamefont {\ifmmode~\dot{Z}\else
  \.{Z}\fi{}ebrowski}},\ }\href {https://doi.org/10.1103/PhysRevB.97.165303}
  {\bibfield  {journal} {\bibinfo  {journal} {Phys. Rev. B}\ }\textbf {\bibinfo
  {volume} {97}},\ \bibinfo {pages} {165303} (\bibinfo {year}
  {2018}{\natexlab{b}})}\BibitemShut {NoStop}%
\bibitem [{\citenamefont {Rostami}\ \emph {et~al.}(2016)\citenamefont
  {Rostami}, \citenamefont {Asgari},\ and\ \citenamefont
  {Guinea}}]{Rostami2016}%
  \BibitemOpen
  \bibfield  {author} {\bibinfo {author} {\bibfnamefont {H.}~\bibnamefont
  {Rostami}}, \bibinfo {author} {\bibfnamefont {R.}~\bibnamefont {Asgari}},\
  and\ \bibinfo {author} {\bibfnamefont {F.}~\bibnamefont {Guinea}},\ }\href
  {https://doi.org/10.1088/0953-8984/28/49/495001} {\bibfield  {journal}
  {\bibinfo  {journal} {Journal of Physics: Condensed Matter}\ }\textbf
  {\bibinfo {volume} {28}},\ \bibinfo {pages} {495001} (\bibinfo {year}
  {2016})}\BibitemShut {NoStop}%
\bibitem [{\citenamefont {Berry}(1984)}]{Berry}%
  \BibitemOpen
  \bibfield  {author} {\bibinfo {author} {\bibfnamefont {M.~V.}\ \bibnamefont
  {Berry}},\ }\href {https://doi.org/10.1098/rspa.1984.0023} {\bibfield
  {journal} {\bibinfo  {journal} {Proceedings of the Royal Society of London.
  A. Mathematical and Physical Sciences}\ }\textbf {\bibinfo {volume} {392}},\
  \bibinfo {pages} {45} (\bibinfo {year} {1984})}\BibitemShut {NoStop}%
\bibitem [{\citenamefont {Wilczek}\ and\ \citenamefont {Zee}(1984)}]{Wilczek}%
  \BibitemOpen
  \bibfield  {author} {\bibinfo {author} {\bibfnamefont {F.}~\bibnamefont
  {Wilczek}}\ and\ \bibinfo {author} {\bibfnamefont {A.}~\bibnamefont {Zee}},\
  }\href {https://doi.org/10.1103/PhysRevLett.52.2111} {\bibfield  {journal}
  {\bibinfo  {journal} {Phys. Rev. Lett.}\ }\textbf {\bibinfo {volume} {52}},\
  \bibinfo {pages} {2111} (\bibinfo {year} {1984})}\BibitemShut {NoStop}%
\bibitem [{\citenamefont {San-Jose}\ \emph {et~al.}(2007)\citenamefont
  {San-Jose}, \citenamefont {Schön}, \citenamefont {Shnirman},\ and\
  \citenamefont {Zarand}}]{SanJose}%
  \BibitemOpen
  \bibfield  {author} {\bibinfo {author} {\bibfnamefont {P.}~\bibnamefont
  {San-Jose}}, \bibinfo {author} {\bibfnamefont {G.}~\bibnamefont {Schön}},
  \bibinfo {author} {\bibfnamefont {A.}~\bibnamefont {Shnirman}},\ and\
  \bibinfo {author} {\bibfnamefont {G.}~\bibnamefont {Zarand}},\ }\href
  {https://doi.org/https://doi.org/10.1016/j.physe.2007.05.027} {\bibfield
  {journal} {\bibinfo  {journal} {Physica E: Low-dimensional Systems and
  Nanostructures}\ }\textbf {\bibinfo {volume} {40}},\ \bibinfo {pages} {76}
  (\bibinfo {year} {2007})}\BibitemShut {NoStop}%
\bibitem [{\citenamefont {Prabhakar}\ \emph {et~al.}(2014)\citenamefont
  {Prabhakar}, \citenamefont {Melnik},\ and\ \citenamefont
  {Bonilla}}]{Prabhakar}%
  \BibitemOpen
  \bibfield  {author} {\bibinfo {author} {\bibfnamefont {S.}~\bibnamefont
  {Prabhakar}}, \bibinfo {author} {\bibfnamefont {R.}~\bibnamefont {Melnik}},\
  and\ \bibinfo {author} {\bibfnamefont {L.~L.}\ \bibnamefont {Bonilla}},\
  }\href {https://doi.org/10.1103/PhysRevB.89.245310} {\bibfield  {journal}
  {\bibinfo  {journal} {Phys. Rev. B}\ }\textbf {\bibinfo {volume} {89}},\
  \bibinfo {pages} {245310} (\bibinfo {year} {2014})}\BibitemShut {NoStop}%
\bibitem [{\citenamefont {Bednarek}\ \emph {et~al.}(2012)\citenamefont
  {Bednarek}, \citenamefont {Pawlowski},\ and\ \citenamefont
  {Skubis}}]{Pawlowski2012}%
  \BibitemOpen
  \bibfield  {author} {\bibinfo {author} {\bibfnamefont {S.}~\bibnamefont
  {Bednarek}}, \bibinfo {author} {\bibfnamefont {J.}~\bibnamefont
  {Pawlowski}},\ and\ \bibinfo {author} {\bibfnamefont {A.}~\bibnamefont
  {Skubis}},\ }\href {https://doi.org/10.1063/1.4714771} {\bibfield  {journal}
  {\bibinfo  {journal} {Applied Physics Letters}\ }\textbf {\bibinfo {volume}
  {100}},\ \bibinfo {pages} {203103} (\bibinfo {year} {2012})}\BibitemShut
  {NoStop}%
\bibitem [{\citenamefont {Devakul}\ \emph {et~al.}(2021)\citenamefont
  {Devakul}, \citenamefont {Cr{\'e}pel}, \citenamefont {Zhang},\ and\
  \citenamefont {Fu}}]{twisted_tmd}%
  \BibitemOpen
  \bibfield  {author} {\bibinfo {author} {\bibfnamefont {T.}~\bibnamefont
  {Devakul}}, \bibinfo {author} {\bibfnamefont {V.}~\bibnamefont {Cr{\'e}pel}},
  \bibinfo {author} {\bibfnamefont {Y.}~\bibnamefont {Zhang}},\ and\ \bibinfo
  {author} {\bibfnamefont {L.}~\bibnamefont {Fu}},\ }\href
  {https://doi.org/10.1038/s41467-021-27042-9} {\bibfield  {journal} {\bibinfo
  {journal} {Nature Communications}\ }\textbf {\bibinfo {volume} {12}},\
  \bibinfo {pages} {6730} (\bibinfo {year} {2021})}\BibitemShut {NoStop}%
\end{thebibliography}%
\end{document}